\documentclass[twocolumn,5p,authoryear]{elsarticle}

\usepackage{graphicx}
\usepackage{natbib}
\usepackage{amsmath,amssymb,latexsym}
\usepackage{multirow}
\usepackage{lineno}
\usepackage{color}
\definecolor{lila}{rgb}{0.3,0,0.7}
\definecolor{grey}{rgb}{0.5,0.5,0.5}

\newcommand{\ME}{M_{\oplus}}
\newcommand{\Tsurf}{\mbox{ $T_{\rm 1\,bar}$ }}
\newcommand{\Ptrans}{\mbox{$P_{\mbox{\tiny1-2}}$}}
\newcommand{\Msat}{\mbox{$M_{\rm Sat}$}}
\newcommand{\Rsat}{\mbox{$R_{\rm Sat}$}}
\newcommand{\Req}{\mbox{$R_{\rm eq}$}}
\newcommand{\Rmean}{\mbox{$R_{\rm m}$}}
\newcommand{\Mcore}{\mbox{$M_{\rm core}$}}
\newcommand{\Rcore}{\mbox{$R_{\rm core}$}}
\newcommand{\Yatm}{\mbox{$Y_{\rm atm}$}}
\newcommand{\Zatm}{\mbox{$Z_{\rm atm}$}}
\newcommand{\Zsol}{\mbox{$Z_{\odot}$}}
\newcommand{\HeH}{\mbox{He:H$_2$}}

\journal{Icarus}

\begin{document}

\begin{frontmatter}


\title{Saturn layered structure and homogeneous evolution models with different EOSs}

\author[UR,UCSC]{Nadine Nettelmann\corref{corA}}\ead{nadine.nettelmann@uni-rostock.de}
\author[UR]{Robert P{\"u}stow}\ead{robert.puestow@uni-rostock.de}
\author[UR]{Ronald Redmer}\ead{ronald.redmer@uni.rostock.de}
\address[UR]{Institute of Physics, University of Rostock, D-18051 Rostock, Germany}
\address[UCSC]{University of California, Santa Cruz, USA\\\vspace{0.2cm}
\footnotesize\textit{Accepted to Icarus, April 2013}}

\cortext[corA]{Corresponding author}

\begin{abstract}
The core mass of Saturn is commonly assumed to be 10--25 $\ME$ as predicted by interior models 
with various equations of state (EOSs) and the \emph{Voyager} gravity data, and hence larger than 
that of Jupiter (0--$10\:\ME$). We here re-analyze Saturn's internal structure and evolution by using 
more recent gravity data from the \emph{Cassini} mission and different physical equations of state: 
the ab initio LM-REOS which is rather soft in Saturn's outer regions but stiff at high pressures, 
the standard Sesame-EOS which shows the opposite behavior, and the commonly used SCvH-i EOS.
For all three EOS we find similar core mass ranges, i.e.~of 0--$20\:\ME$ for SCvH-i and Sesame EOS  
and of 0--$17\:\ME$ for LM-REOS.
Assuming an atmospheric helium mass abundance of 18\%, we find maximum atmospheric metallicities, 
{\Zatm} of $7\times$ solar for SCvH-i and Sesame-based models and a total mass of heavy elements, 
$M_Z$ of 25--$30\ME$. Some models are Jupiter-like. With LM-REOS, we find 
$M_Z=16$--$20\ME$, less than for Jupiter, and $\Zatm \lesssim 3\times$ solar. For Saturn, we compute
moment of inertia values $\lambda=0.2355(5)$. Furthermore, we confirm that homogeneous evolution 
leads to cooling times of only $\sim 2.5$ Gyr, independent on  the applied EOS. 
Our results demonstrate the need for accurately measured atmospheric helium and oxygen abundances, 
and of the moment of inertia for a better understanding of Saturn's structure and evolution.
\end{abstract}


\end{frontmatter}


\section{Introduction}

Saturn is the planet with the lowest mean density in the solar system. Since 
the mechanisms that can inflate exoplanets with observed overlarge radii do not hold for the outer
planet Saturn, one might thus  intuitively think of Saturn as having a smaller core and smaller 
overall metallicity than Jupiter. However, quantitative estimates on the core mass and on the
total heavy element enrichment solely come from interior model calculations, and the same modeling 
approach applied to both planets just predicts the opposite: an about two times 
larger maximum core mass and heavy element enrichment for Saturn \citep{SG04,GuiGau07}.  A higher 
envelope metallicity of Saturn is also supported by the measured atmospheric C:H ratios, which is 
$\sim 9\times$ solar for Saturn (\citealp{Fletcher+09}; scaled to the solar system abundance data 
of \citealt{Lodders03}) but only 3--5$\times$ solar for Jupiter \citep{Atreya+03}. 

Certainty about the present core mass and envelope metallicity is desirable because these parameters 
contain information ---albeit not necessarily uniquely \citep{Helled+10,Boley+11}--- on the formation 
environment, i.e.~on the protosolar disk, and on the process of formation. 

Models by \citet{SG04}, hereafter SG04,  are often considered the standard of what we know today 
about Saturn's present internal structure in terms of core mass and heavy element enrichment 
(\citealp[e.g.,][]{Alibert+05,DodRob+08}), for mainly two reasons. First, these models have been 
computed for various  physical equations of state(EOS) for Saturn's likely 
main constituents H and He that also give acceptable solution for Jupiter's interior and evolution 
(the EOSs SCvH-i, LM-H4, LM-SOCP). Independent on the EOS, the possible core mass range was found to be 
$\sim 10$--$25\ME$, while for Jupiter $\sim 0$--$10\ME$.
Second, a wide range of input parameters was accounted for such as a the position of an 
internal layer boundary that separates a helium-poor, outer from a helium-rich, inner envelope. 
However, SG04 computed constant metallicity envelope models only, an assumption that 
tremendously restricts the resulting range of interior models.

In earlier models by Gudkova \& Zharkov (1999) and \citet{Gui99}, the metallicity was allowed to vary
across the internal layer boundary. As a consequence, zero-core mass models with high heavy
element enrichment in the deep envelope were found for both Jupiter and Saturn.  

The new \emph{Cassini} gravity data with their tight observational error bars, and also long-term 
observational data of the Saturnian system \citep{J+06, AS07} raised hope to better constrain Saturn's 
internal structure. Surprisingly, the most recent Saturn models based on those gravity data cover an 
even bigger, minimum core mass range of $\sim 0$--$30\ME$ \citep{AS07,Helled+09a,Helled11}. Therefore, 
\cite{Helled11} suggests to measure the axial moment of inertia as an additional constraint. Her models,
however, employ \emph{empirical} pressure-density relations that may reach out of the realm of physical 
EOS which agree with the available experimental data  (see, e.g.~SG04; \citealt{Holst+12}).

Our Saturn models are the first that are based on both physical equations of state and the \emph{Cassini} data. 
Not is it the purpose of this work to better constrain the core mass: this cannot be achieved within the 
standard three-layer modeling approach, which is adopted in this work. Instead, we here investigate the 
overall behavior of core mass, atmospheric metallicity, and deep envelope metallicity on the input parameters: 
we vary the position of an internal layer boundary in order to recall its influence on the core mass, 
see also \citet{GuiGau07}; we exchange the EOS of the envelope material 
(LM-REOS, SCvH-i EOS, Sesame EOS), and we adopt two different periods of rotation of 10h 32m and 10h 39m.  
In lack of accurate observations, we make predictions on the possible helium and heavy element mass 
fractions in Saturn's atmosphere in dependence on the $J_4$ value and the uncertainty in the 
rotational period. 
Our results on the atmospheric helium abundance can serve as constraints for future models of 
He-sedimentation in Saturn, as long as  Saturn's atmospheric He:H$_2$ ratio is not accurately measured.

Observations of young stellar systems and protostellar disks commonly point to formation of the giant 
planets within a few Myr \citep{Strom+93}, implying a billions-of-years-old planet should have the 
same age as its host star.
However, homogeneous evolution calculations for Saturn, which are mainly based on the SCvH-i EOS, 
generally yield cooling times of 2--3~Gyr \citep{Saumon+92,Fortney+11}, about only
\emph{half} of the age of the Sun. This implies a higher luminosity of present Saturn than it 
should have if the underlying assumption of homogeneous evolution would hold. Despite the obvious failure 
of this assumption, we here adopt it once more in order to investigate the influence of the EOS
on the cooling time.

In Section \ref{sec:m_modeling} we describe our modeling procedure. Section~\ref{sec:m_obs} 
is devoted to a detailed description of the observational data, and Section \ref{sec:m_eos} 
to the applied EOSs. Our results are presented in Section \ref{sec:results}. In Section 
\ref{sec:r_struc1} we investigate the influence on different H-He-EOS on Saturn's structure and in 
Section \ref{sec:r_struc2} of the atmospheric He abundance and rotation rate. 
In Section \ref{sec:r_nmoi} we give the values for the non-dimensional moment of inertia. Section 
\ref{sec:r_evol} contains the cooling curves. Section~\ref{sec:discussion} includes a discussion 
on the implications for Saturn's formation process (\ref{sec:d_mcoreform}), on the applicability 
of the three-layer assumption in the presence of He rain (\ref{sec:d_Herain}), and a summary 
of our main findings (\ref{sec:d_summary}).

\section{Methods}\label{sec:methods}

\subsection{Planetary structure modeling}\label{sec:m_modeling}

For understanding the interior of giant gas planets like Saturn it is necessary 
to consider the gravitational field of the planet. The shape of the field is influenced by different effects. 
Saturn for instance has primarily the form of an ellipsoid due to its rapid rotation, which can be seen 
from the rather high ratio of centrifugal to gravitational forces, $q=\omega^2\,R^3_{\rm eq}/(G\,M)$,
where $\omega$ is the angular velocity, $R_{\rm eq}$ is the equatorial radius, and $M$ the total mass.
For Saturn, $q\sim 0.155$ with an uncertainty of 0.004 due to the uncertainty in the 
rotation period and equatorial radius (see Section \ref{sec:m_obs}), for Jupiter, $q=0.089$, and for the Sun, 
$q=0.00002$. Tidal forces caused by the gravity of the moons or the parent star can also change the form 
of a planet's gravity field. While this effect can be important for close-in exoplanets it is tiny for 
Saturn and has not been measured yet for any giant planet in the solar system. 
To assess the rotationally induced deformation, the gravity field $\Phi^{(e)}$ exterior to the 
mass $M$ is expanded into a series of Legendre polynomials $P_{2n}$,  
where the expansion coefficients $J_{2n}$ are the gravitational moments at the
equatorial reference radius $\Req$,
\begin{equation}\label{eq:J2n}
	J_{2n}=-\frac{1}{MR_{\rm eq}^{2n}}\int\! d^3r\,\rho(r,\theta)r^{2n}P_{2n}(t)\:.
\end{equation}
Being integrals of the internal mass distribution over the volume enclosed 
within the geoid of equatorial radius $\Req$, the $J_{2n}$ can be written 
as depth-dependent functions $J_{2n}(l)$ whose values increase continuously from the center outward until
the observed values $J_{2n}^{(\rm obs)}$ are reached at the geoid's mean radius $l=\Rmean$. As a measure 
for the contribution $dJ_{2n}$ of a shell at $l$ and extension $dl$ to $J_{2n}^{(\rm obs)}$ 
we can define the normalized contribution function 
\begin{equation}\label{eq:c2n}
	c_{2n}(l) = \frac{(dJ_{2n}/dl)|_{l}}{\int \! dl'\, (dJ_{2n}/dl')}\:.
\end{equation}
For modeling Saturn we use the same method and code as in \citet{N+12} for Jupiter. 
We adopt the standard three-layer structure with two envelopes and a core. The composition of each of 
the envelopes is diverted into the three components hydrogen, helium, and heavy elements, whereas 
the core consists of heavy elements only.
The helium mass fractions and the metallicities (i.e.~the heavy element mass fractions) are parameterized 
by $Y_1$, $Z_1$ and $Y_2$, $Z_2$ for the outer and the inner envelope, respectively. This implies the
assumption of homogeneous envelopes. The transition between them occurs at the transition pressure 
$\Ptrans$ which is a free parameter. 
As observational constraints we take into account $\Req$, $\omega$, the total mass \Msat, the temperature 
$T_1$ at the 1~bar level of the planet, and the lowest order moments $J_2$ and $J_4$. 

For given values of $Y_1$ and of the mean helium abundance $Y$, $Y_2$ is adjusted 
to fit $Y$, while $Z_1$ and $Z_2$ are adjusted to fit $J_2$ and $J_4$. Mass conservation 
is then ensured by the choice of the core mass $\Mcore=m(\Rcore)$.

\subsection{Observational constraints}\label{sec:m_obs}

While the \emph{Cassini} mission could provide tight constraints on Saturn's gravity field, 
there are still important remaining uncertainties, in particular in Saturn's period of rotation, 
equatorial radius, and the atmospheric helium abundance.

\paragraph{Period of rotation}

Prior to the \emph{Cassini} observations, Saturn's period of rotation was taken to be 10h 39m 24s, the detected
periodicity in the kilometric radio emissions of Saturn's magnetic field as measured by the Voyager I 
and II spacecraft \citep{DK81}. \emph{Cassini} however revealed a prolongation of this period by several 
minutes within just 20 years; thus the observed magnetic field modulations may not reflect the rotation 
of Saturn's deep interior \citep{Gurnett+07}. 
On the other hand, while alternative methods of deriving the rotation rate from observed wind speeds make 
assumptions that may not hold true, such as the minimum energy of the zonal winds or a minimum height of 
isobar-surfaces relative to computed geoid surfaces \citep{AS07,Helled+09b}, that alternative methods just 
suggest similar values of $\sim$ 10h 32m. We therefore use these values as the uncertainty in Saturn's 
real solid body rotation period and compute interior models for both periods, i.e.~for 10h 32m and 10h 39m.
Note that we neglect here the uncertainty to Saturn's structure from the possibility of differential 
rotation on cylinders. On the other hand, all observational wind data can well
be reproduced by the assumption of solid-body rotation \citep{Helled+09b} and the effect of zonal winds on 
Jupiter's structure have been shown to be negligible if their penetration depth is limited to 1000 km 
\citep{Hubbard99}.

\paragraph{Equatorial radius}

Our method of interior modeling requires all outer boundary conditions to be provided at the 
same pressure level. Although the Voyager~I,~II, and Pioneer~11 observational data refer to the 
isobar-surface at 100~mbar \citep{Lindal+85}, we prefer to use the 1~bar level as outer boundary. 
In particular, we use the equatorial 1~bar radius of $R_{\rm eq}=60,268$~km \citep{Lindal+85, GuiGau07}. 
This radius is a computed one of a geoid that rotates 
with a solid body period (System III period) of 10h 39m 24s plus an additional, latitude~($\phi$)-dependent 
component according to the observed zonal wind speeds. The difference in radius at the equator
to a reference geoid for a rigidly rotating Saturn in hydrostatic equilibrium ---the appropriate 
radius to constrain interior models--- is $\sim 120$~km at the 100~mbar level. The exact difference 
at the 1~bar level is not provided (see \citealp{Lindal+85} for details), but we expect it to be
of same size, and therefore would expect the $\Req$ value of a reference geoid at 1~bar to 
be $\sim 100$~km smaller than the given value in \citet{Lindal+85}.
Neglecting the described inconsistency, we here use this $\Req$ value, and we use it 
independently on the period of rotation value. To our awareness, the only consistent reference systems 
for both limiting periods of rotation (10h 39m 24s and 10h 32m 35s) are presented in 
\citet{Helled11}. Her reference system values are listed in Table \ref{tab:obs} for comparison.
The surface radius $\Rsat$ is the mean radius of the reference geoid with the equatorial radius $R_{\rm eq}$.

\paragraph{Gravitational coefficients}

In order to get the gravity field ($J_2$, $J_4$, $J_6$) at the outer boundary as defined 
by the equatorial 1~bar radius of 60,268~km, we scale the values of \citet{J+06}, which are 
$J'_2=16290.71(0.27)$, $J'_4=-935.8(2.8)$, and  $J'_6=86.1(9.6)$. We use the scaling relation 
$J_{2n}' \Req^{'\,2n} = J_{2n} \Req^{\,2n}$, where $\Req^{'\,2n}=60,330$~km is the equatorial 
reference radius in \citet{J+06}. The gravitational coefficients of \citet{J+06} are based on, 
but not limited to, Pioneer11, Voyager, and \emph{Cassini} tracking data as well as long-term Earth-based 
and HST astrometry data. They have significantly reduced error bars compared to the Voyager era data 
\citep{Campbell+89}. Since the $\Req$ value used in this work is larger than that in \citet{Helled11}, 
the scaled, absolute  $J_{2n}$ values are smaller. Under variation of the $J_4$ value within its $6\:\sigma$ 
error bars (Section \ref{sec:r_struc2}), we also cover the $J_4$ values used in \citet{Helled11}, 
see Table \ref{tab:obs}. Therefore, the obtained sets of solutions can reasonably be compared to each other.

\paragraph{Mean helium abundance}

For the protosolar cloud where the Sun and the giant planets formed of, Bahcall \& Pinsonneault~(1995) 
calculate a mean helium abundance of 27.0 to 27.8\% by mass, depending mainly on the inclusion of helium 
and heavy element diffusion into solar evolution models. We require our models to have and mean helium 
abundance $Y=0.2750(1)$ by mass with respect to the H/He subsystem. By particle numbers, this corresponds 
to He:H$_2\::=\:N_{\rm He}/N_{\rm H_2} = 0.1732$.

\paragraph{Atmospheric helium abundance}

Combined \emph{Voyager} infrared spectrometer (IRIS) data and Voyager radio occultation (RSS) temperature 
profile data  revealed a depletion of He in the atmospheres of both Jupiter and Saturn compared to the 
protosolar value \citep{Conrath+84}.
In particular, a modest depletion He:H$_2=0.110\pm 0.032$ was found in Jupiter, and a strong depletion 
He:H$_2=0.034\pm 0.024$ in Saturn. These particle ratios correspond to atmospheric mass mixing ratios 
$Y_{\rm atm,\, J}=0.18\pm 0.04$ for Jupiter and $Y_{\rm atm,\, S }=0.06\pm 0.05$ for Saturn. However, 
the He abundance detector (HAD) aboard the Galileo probe measured \emph{in situ} Jupiter's atmospheric 
He:H$_2$ to be $0.157\pm 0.003$ \citep{Zahn+98}, corresponding to $Y_{\rm atm,\, J}=0.238 \pm 0.006$. 
Because of the discrepancy to the former results, the Voyager data for Jupiter and Saturn were re-examined 
by \citet{CG00}. They found the Voyager data for Jupiter  could be made consistent with the Galileo data 
if the temperature profile obtained by Voyager RSS measurements were shifted by 2~K towards colder 
temperatures. Because of a possible systematic error of the RSS data, \citet{CG00} developed 
an inversion algorithm to infer the He:H ratio and the temperature profile from the Saturnian IRIS spectra 
alone. Their results suggest a significantly larger He abundance $Y_{\rm atm,\, S}=0.18$--0.25,
in agreement with the early Pioneer IR data based determination of $0.182\pm 0.005$ (He:H$_2=0.111\pm 3\%$) 
\citep{OrtonIng80}.  
Unfortunately, the method of \citet{CG00} cannot be checked by application to Jupiter due to disturbing NH$_3$
cloud formation in the spectral range of interest in Jupiter's slightly warmer atmosphere. 

\citet{Kerley04a} instead suggests to trust the ratio of the He abundances for the two planets 
as derived originally from the Voyager IRIS and RSS data. With 
$(\HeH)_{\rm S} = 0.31(1\pm 1.1)\times(\HeH)_{\rm J}$ and $m_{\rm He}=4$~g/mol, $m_{\rm H_2}=2$~g/mol, 
this would give
\begin{equation}
	Y_{\rm atm, \,S} 
  = \frac{(\HeH)_{\rm S}}{(\HeH)_{\rm J}}
	\:\frac{m_{\rm H_2} + m_{\rm He}(\HeH)_{\rm J}}{m_{\rm H_2} + m_{\rm He}(\HeH)_{\rm S}} \:Y_{\rm atm,\, J}\:,
\end{equation}
which is in numbers
\begin{equation}\label{eq:Y_Kerley}
	Y_{\rm atm, \,S}	= 0.353\,(1\pm 2)\times 0.238 = 0.084\,(1\pm 2)\:.
\end{equation}
We thus compute Saturn models for different helium abundances $Y_{\rm atm}$ between 0.10 and 0.18. 
Because of the assumed convection below the 1~bar level, $Y_1 = Y_{\rm atm}$.

\paragraph{Atmospheric temperature}

Saturn's atmospheric temperature was determined to be $135\pm 4$~K at the 1~bar level and
$145\pm 4$~K at 1.3~bars \citep{Lindal+85}. For our model calculations we assume a slightly higher 
1~bar temperature \Tsurf of 140~K but also compute single models with $\Tsurf=135$~K. Note that
the physical temperature is not a direct observable. In particular, the \emph{observational constraint}
\Tsurf depends on an assumed composition \citep{Lindal+85}. It is the height-dependent refractivity 
which was measured during egress and ingress of the Voyager II spacecraft. From these data and 
the assumed composition, the mass density and thus the particle number density can be derived. 
Integration of the equation of hydrostatic equilibrium, $dP/dh=-g\,\rho$, over height $h$ in 
the atmosphere  allows to relate density to pressure. Finally, the thermal equation of state 
for the assumed composition with mean molecular weight $\bar{\mu}$ yields the temperature 
$T(P,\rho)$, which is proportional to $\bar{\mu}P/\rho$. Since the above cited \Tsurf value 
was derived for the low $Y$ value of 0.06 \citep{Lindal+85}, a revision of that value may also 
require a revision of the atmospheric temperature determination towards higher temperatures. 
Table \ref{tab:obs} summarizes the used constraints.

\begin{table}
\caption{\label{tab:obs}Observational constraints for Saturn}
\begin{tabular}{lcccc}
\hline\hline
Parameter & default value & \multicolumn{2}{c}{Helled (2011)}\\
\hline
$P=2\pi/\omega$ & 10h 39m & 10h 39m 24s$^{a}$, & 10h 32m 35s$^{c}$ \\
 & 10h 32m \\
$\Msat$ (10$^{29}$g) & $5.683566^{d}$ & -- & --\\
$R_{eq}$ (km) &  60268$^{b}$ & 60141.4 & 60256.9\\
$J_2\:(10^{-4})$ & 16324.2(0.3) & 16393.1 & 16330.2  \\
$J_4\:(10^{-4})$ & -939.6(2.8) & -947.6 & -940.4\\
$J_6\:(10^{-4})$ & 86.6(9.6) & 87.8 & 86.8\\
$T_{1\,\rm bar}$ (K) & 140~K & -- & --\\ 
$T_{\rm eff}$ (K) & $95.0\pm 0.5^{e}$ & -- & --\\ 
\hline
\end{tabular}
\begin{minipage}{0.5\textwidth}\footnotesize
${}^{a}$ \cite{DK81}; from kilometric radiation and magnetic field data\\
${}^{b}$ \cite{Lindal+85}; for $P$=10h 39m 24s \\
${}^{c}$ \cite{AS07}; 100mbar isosurface height minimization\\
${}^{d}$ with $G\Msat$ from \citet{J+06} and $G=6.67384\:10^{-14}\:\rm m^3\,s^{-2}$ (CODATA 2010)\\
${}^{e}$ \cite{GuiGau07}
\end{minipage}

\end{table}

\subsection{Planetary evolution modeling}

We compute the cooling time of Saturn almost exactly as in \citet{N+12} for Jupiter.
In particular, out of the set of possible structure models for present Saturn, we pick one model,
implying known values of the structure parameters $\Mcore$, $Y_2$, $Z_1$, and $Z_2$.
Keeping these parameter values constant, we then generate a series of models with increased \Tsurf 
values. For these models, we also keep the angular momentum, $L$ conserved and store the energy of 
rotation, $E_{\rm rot}=L\omega /2$. 
The higher $\Tsurf$, the warmer the interior and hence the larger the planet radius $R_p$ and the 
higher the luminosity. Finally, \Tsurf is mapped  onto time by integrating the
cooling equation
\begin{equation}\label{eq:cool}
	dt = -\frac{\int_{\Mcore}^{\Msat}dm\:T(t)\,ds(t) + \Mcore c_v dT_{\rm core}(t) + dE_{\rm rot}(t)}
           {4\pi R^2(t)\,\sigma\, (T_{\rm eff}^4(t)-T_{\rm eq}^4(t)) - L_{\rm radio}(t)}\quad. 
\end{equation}
from present time $\tau_{\odot}=4.56$~Gyr backwards.
In Equation (\ref{eq:cool}), parameters that change with time are written as a function of time,
although most of them do not depend on time explicitly; in fact, only the solar luminosity
$L_{\odot}$, which is proportional to $T_{\rm eq}^4$ and assumed to increase linearly with time,   
and the luminosity $L_{\rm core}$ from the decay of radioactive elements in the core do so. 
Unlike in \citet{N+12} where Jupiter's $L_{\rm radio}$ was kept constant, we here include its time-dependence 
as in \citet{N+11}, because Saturn's $\Mcore/M$ ratio can be $\sim 6$ times larger than Jupiter's and 
thus special core contributions potentially be non-negligible.
The relation between $T_{\rm eff}$  and \Tsurf is adapted from \citet{Guillot+95} with constant 
$K=1.565$. As \Tsurf  determines $R_p$ and $T_{\rm eff}$, the map $\Tsurf\mapsto t$ includes the 
more familiar maps $R_p\mapsto t$ and $T_{\rm eff}\mapsto t$.
We then repeat the described procedure to compute the cooling curve for another structure model of 
present Saturn in order to learn about, e.g., the effect of the chosen EOS on the cooling time.

\subsection{Equations of state}\label{sec:m_eos}
  
Saturn's mantle is believed to mainly consist of hydrogen and helium, and some amount of heavier 
atoms or molecules which we call \emph{heavy elements}.  For each of these three components we 
apply a separate EOS. To obtain an EOS for the mantle material, we linearly mix the single-component 
EOS. Saturn models are then computed for different sets of equations of state for Saturn's mantle: 
LM-REOS (see \citealp{N+08,N+12} and references therein), SCvH EOS for hydrogen and helium by \citet{SCvH95} 
where we mimic heavy elements by scaling the density of the He EOS by a factor of 3/2, and the Sesame EOS. 

The Sesame-5251 hydrogen EOS \citep{SESAME} is the deuterium EOS 5263 scaled in density as developed 
by Kerley in 1972. It is based on the chemical picture and  built upon the assumption of three phases: 
a molecular solid phase, an atomic solid phase, and a fluid phase which takes into account chemical 
equilibrium between molecules and atoms and ionization equilibrium between atoms, protons and electrons.
A completely revised version \citep{Kerley03} includes fits to more recent shock compression data 
resulting into a larger compressibility at $\sim 0.5\:$Mbar, and a smaller one at $\sim 10\:$Mbar. 
We here apply the earlier version as it shows stronger deviations to the SCvH EOS and the LM-R EOS, 
which allows to attribute differences in the resulting Saturn models more clearly to properties of the EOS.
For application to planetary models, the Sesame H EOS is linearly mixed with the helium EOS of 
\citet{Kerley04b} and with the water EOS H$_2$O-REOS.

\subsubsection{Core material}

The cores of giant planets are usually assumed to consist of ices and rocks as a relict from their
time of formation \citep{Mizuno80,Miller+11}. As limiting cases, we either compute models with pure 
water cores using the H$_2$O-REOS \citep{N+08,French+09}, or with pure rocky cores using the 
$P-\rho$ relation of \citet{HM89}. This approach ignores the possibility of an eroded core 
that would also contain some hydrogen and helium.

\subsubsection{Comparison of the applied hydrogen EOSs}

\begin{figure}
\includegraphics[width=0.48\textwidth]{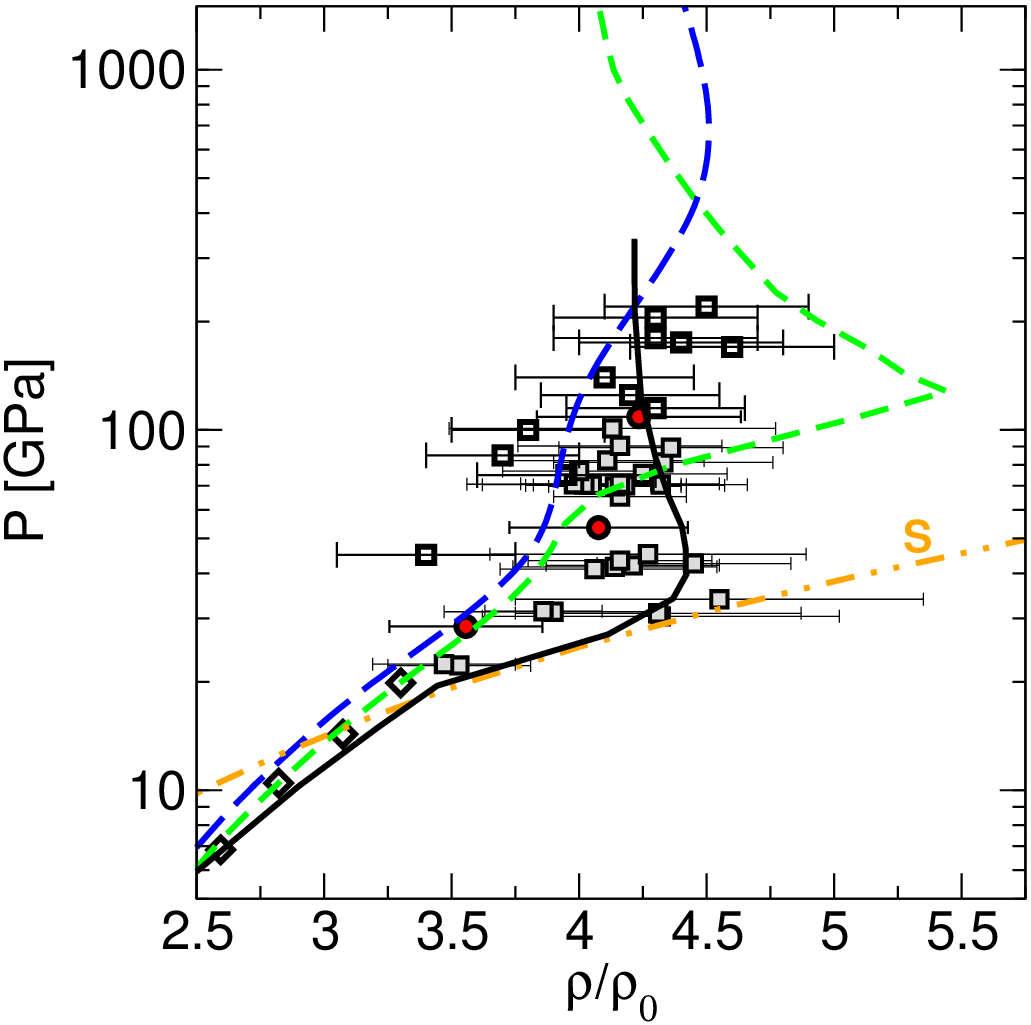}
\caption{\label{fig:hugos} (Color online) 
Theoretical hydrogen Hugoniot curves (\emph{solid, black}: H-REOS.2, \emph{short-dashed}: H-SCvH-i, 
\emph{long-dashed}: Sesame-5251) and experimental shock data (\emph{grey filled squares}:
SNL Z-pinch, \emph{open squares}: modified omega laser, \emph{circles}: explosives, \emph{diamonds}: 
gas gun). The \emph{dot-dot-dashed orange curve (S)} shows part of the Saturn adiabat. }
\end{figure}

Even if the heavy elements in Saturn would have the low molecular weight of helium, still at least 70\% 
of the particles in Saturn would be hydrogen. Therefore, among the equations of state for planetary 
materials, that of hydrogen is expected to have the biggest influence on the resulting structure models. 
In Figure~\ref{fig:hugos} we compare the  above described hydrogen EOSs with experimental deuterium 
shock compression data. For a certain experimental set-up and given initial conditions $(\rho_0, T_0)$, 
different final compression ratios $\rho/\rho_0$ and pressures  can be achieved depending on the velocity 
of the flyer plate upon impact, be it accelerated by a gas gun, magnetic pressure, laser light, or 
explosives. The first-shock end states follow the Hugoniot-relation
\begin{equation}\label{eq:hugo}
	u-u_0 = 0.5(P-P_0)(\rho_0^{-1}-\rho^{-1})\:,
\end{equation}
where initial pressure $P_0(\rho_0,T_0)$ and initial internal energy $u_0(\rho_0,T_0)$ are derived from 
the EOS. Figure \ref{fig:hugos} shows experimental gas gun data from \citet{Nellis+83}, Sandia Z machine 
data from \citet{Knudson+04}, the modified Omega laser data from \citet{KnudDesj09}, and spherical 
compression data using explosives from \citet{Boriskov+05}. By scaling the initial density, theoretical
hydrogen EOSs can reasonably be compared to deuterium experimental data, although some differences 
between D and H become non-negligible in the molecular region, which are probed by the gas gun data,
due to differences in the molecular vibrational states; see \citet{Holst+12} for a detailed 
discussion. Figure \ref{fig:hugos} also shows the theoretical hydrogen Hugoniot curves for H-REOS.2 
(\citealp{Holst+12}, with additional data points by A.~Becker \emph{pers.~comm}), for the H-SCvH-i EOS, 
and for the H-Sesame EOS. 

Obviously, the theoretical hydrogen EOSs differ substantially from each other. Along the Hugoniot states, 
the Sesame EOS, which precedes even the gas gun data, is relatively stiff at $P\lesssim 1$~Mbar but the most
compressible one at $\sim 5$~Mbars. Conversely, the ab initio H EOS is relatively compressible below 1~Mbar
compared to the Sesame and SCvH-i EOS, and to the spherical-shock compression data, with a maximum compressibility 
at $P\sim 0.5$~Mbar where dissociation occurs. At higher pressures of 1 to 3 Mbar, the ab initio EOS runs 
nearly through the experimental central values which indicate a low compression ratio of 4.25, whereas SCvH-i EOS 
agrees well with the data up to 0.8~Mbars but then turns to a large maximum compressibility at $\sim 1$~Mbar 
where in the underlying SCvH-ppt EOS the plasma phase transition occurs. These properties lead to 
systematic differences in the resulting Saturn models.

\section{Results}\label{sec:results}

\subsection{Structure models with different equations of state}\label{sec:r_struc1}

\begin{figure}
\includegraphics[width=0.4\textwidth]{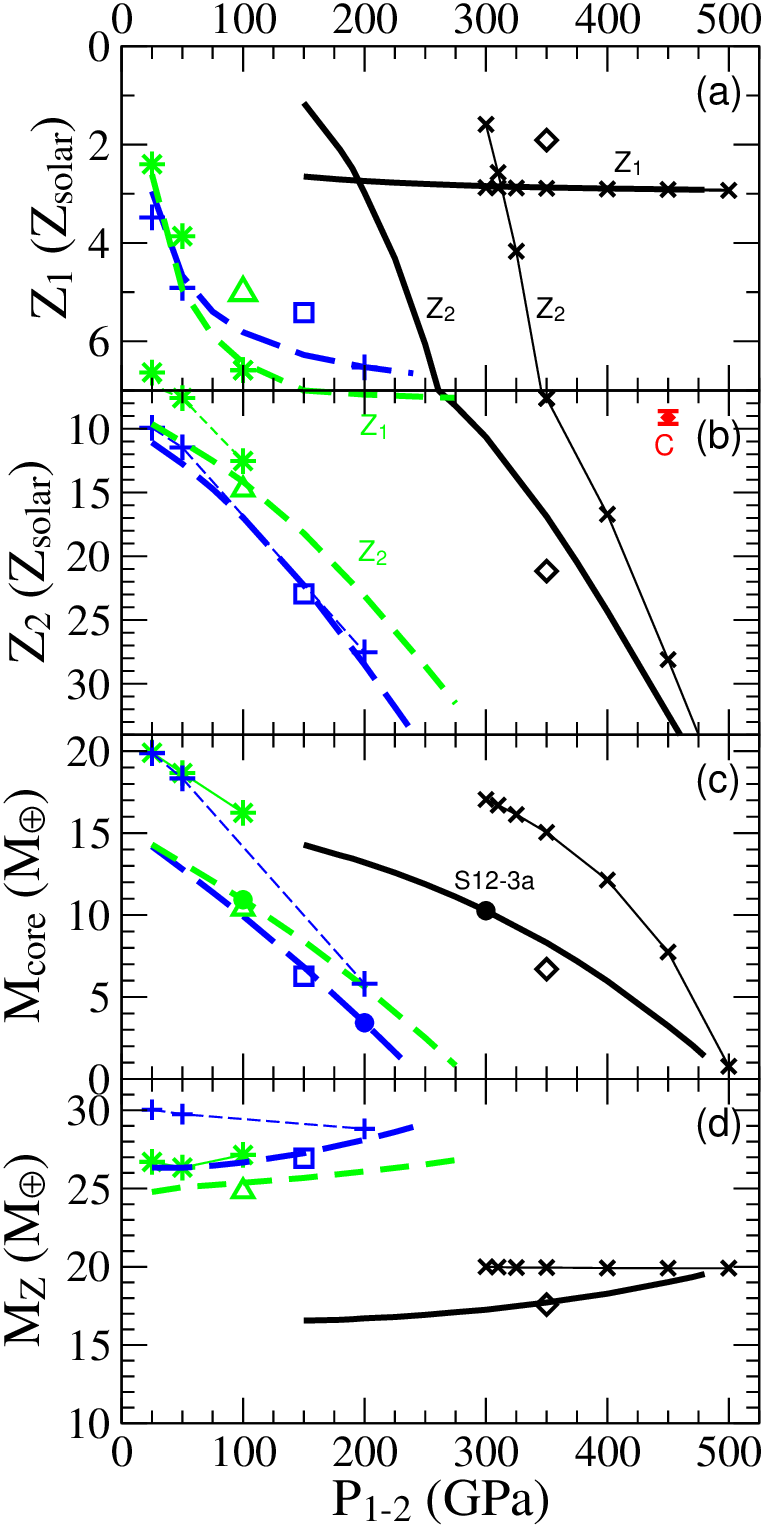}
\caption{\label{fig:mcZZ_P12} (Color online) 
Resulting Saturn structure models for different equations of state  for the mantle (\emph{solid, black}: LM-REOS; 
\emph{long-dashed, blue}: Sesame EOS; \emph{dashed, green}: SCvH-i EOS) and for the core (\emph{thick lines}: 
rocky cores, \emph{symbols connected by thin lines}: water cores), and different surface temperatures 
(\emph{open symbols}: $\Tsurf=135$~K). For given other input parameters (see Section \ref{sec:m_modeling} 
for details), each displayed model is uniquely defined by its $\Ptrans$-value ($x$-axis). Note that the two 
upper panels have a common $y$-axis which changes scale at $Z_1=Z_2=7\times Z_{\rm solar}$,
where $Z_{\rm solar}=1.5\%$. One model is highlighted for each EOS by a filled circle in panel (c). 
The \emph{red diamond} shows the measured $9.12\times$ solar C/H ratio.} 
\end{figure}

In this Section we focus on the effect of the different EOSs (Sesame, SCvH, LM-REOS)
on the resulting Saturn models. The presented models have been calculated for $Y_1=0.18$, 
$2\pi/\omega=$10h~39m, and the default values of the other observational constraints as given 
in Table \ref{tab:obs}. Figure \ref{fig:mcZZ_P12} shows the results.

For each of the considered mantle EOS, the parameters $Z_1$, $Z_2$, $\Mcore$, and $M_Z$ behave similarly 
with $\Ptrans$. The deeper the layer boundary, the higher become $Z_1$ and $Z_2$ and, as a response, 
the lower becomes $\Mcore$, while $M_Z$ remains nearly constant within $\sim 5\ME$. 
Zero-core mass models are possible simply by putting the layer boundary sufficiently 
deep into the planet. The maximum core mass is determined by the condition that neither $Z_1$ nor $Z_2$ 
must become negative. It is $15\ME$ for rocky cores and about $20\ME$ for water cores. 

For given $\Ptrans$, water cores are up to 50\% heavier than rock cores because low-density material 
requires a larger volume. Displaced envelope material must be replaced by core material, hence the core 
becomes heavier.  Note that water core models that approach the limit $\Mcore \to 0$ have been 
computed for LM-REOS mantle EOS only, but they exist for the other mantle EOS as well.

The core mass responds weakly to the surface temperature. The found slight decrease by $\leq 2\ME$ can 
intuitively be explained by the colder and thus denser envelopes. However, the colder outer envelope 
requires $\Delta Z_1\gtrsim 0.01$ less heavy elements to match $J_4$, but up to $\Delta Z_2=0.05$ \emph{more} 
heavy elements in the inner envelope. Thus the found insensitivity of $\Mcore$, and also of $M_Z$, to 
the surface temperature appears to be a more complex compensation of different effects in Saturn. 

\begin{figure}  
\includegraphics[width=0.48\textwidth]{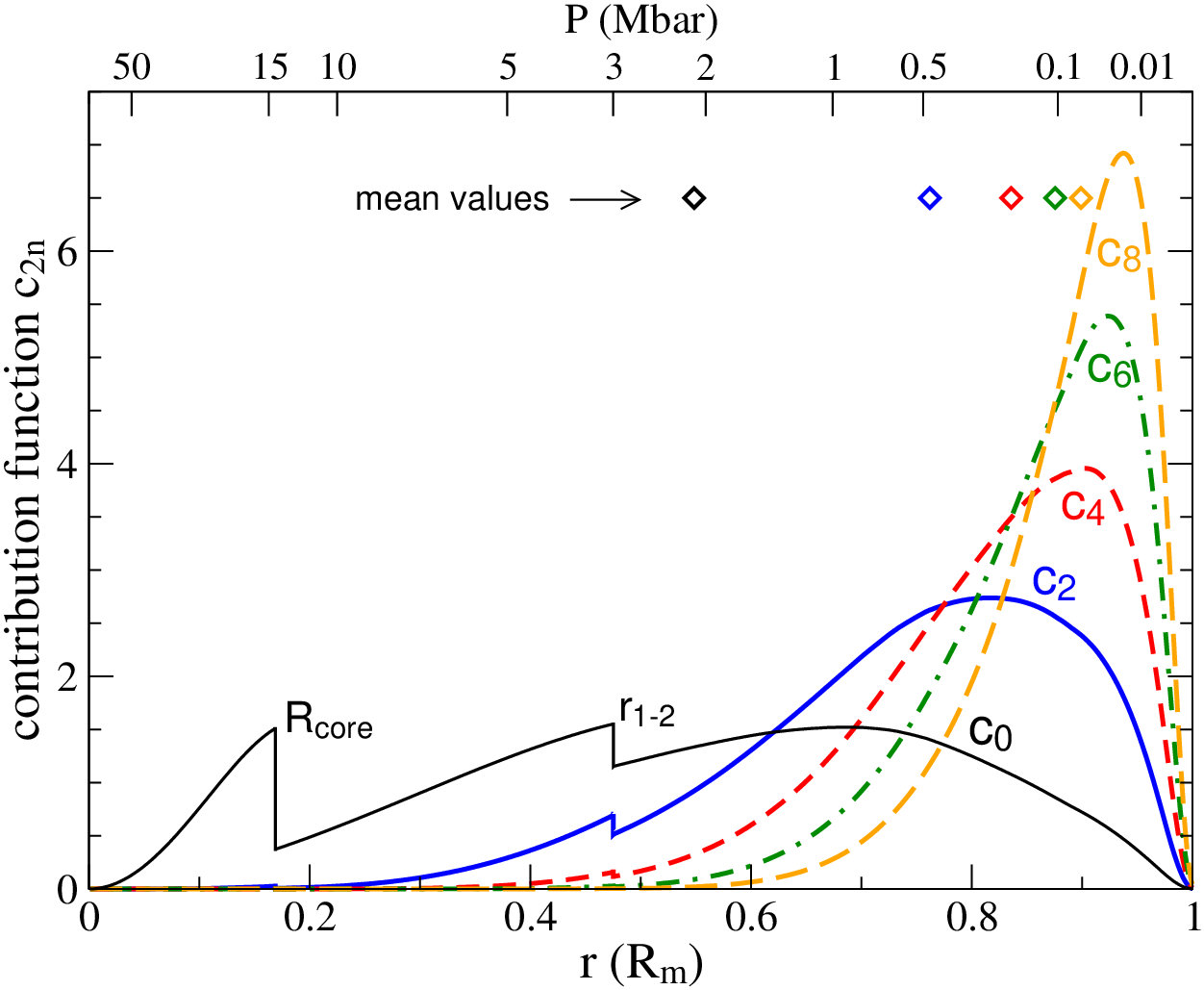}
\caption{\label{fig:contribJ2n} (Color online)
Contribution functions of the gravitational harmonics $J_2$ (\emph{thick solid, blue}), $J_4$ 
(\emph{short-dashed, red}), $J_6$ (\emph{dot-dashed, green}), and $J_8$ (\emph{long-dashed, orange}) 
for the Saturn model S12-3a. The \emph{bottom x-axis} scales linearly with mean radius; 
the \emph{top x-axis} shows the radii where the pressures of 0.01--50~Mbar occur. 
Layer boundaries are clearly seen in the function $c_0$ (\emph{thin solid, black}). 
\emph{Diamonds} show the radius where half of the final $J_{2n}$ value is reached.}
\end{figure}

The general behavior of the solutions can be understood with the help of Figure \ref{fig:contribJ2n}.
It shows the sensitivity of the gravitational moments to the internal mass distribution as parameterized 
by the contribution functions $c_{2n}$, see Eq.~(\ref{eq:c2n}). While they have been computed for a 
particular model, i.e.~the LM-REOS based Saturn model S12-3a as highlighted in Figure \ref{fig:mcZZ_P12}c, 
their properties are the same for all three-layer models. As it is well knowm \citep{ZT78}
the higher the order of the gravitational harmonic, the farther out are the locations of mean and maximum 
sensitivity and the more pronounced is the latter one. $J_2$ and $J_4$ are most sensitive at pressures 
of $\sim 0.5$ and $0.1$~Mbar, respectively. At the 1~Mbar level, the sensitivity of $J_4$ has 
dropped to $\sim 30$\% of its maximum value. $J_4$ is almost insensitive to the mass below 3~Mbars, 
a typical pressure for the outer/inner envelope boundary of LM-REOS based Saturn models. Therefore, 
$Z_1$ changes little with $\Ptrans$ for $\Ptrans > 1$~Mbar.
As the sensitivity of $J_2$ is similar to that of $J_4$ and just slightly shifted to higher pressures,
the computed $J_2$ value is affected by the value of $Z_1$ as well. But because $J_2$ is to be adjusted 
by the $Z_2$ value, i.e.~by the mass distribution interior to $\Ptrans$ where its sensitivity is low, 
small changes in $J_2$ for adjustment require strong changes in $Z_2$, see Figure \ref{fig:mcZZ_P12}b. 

\paragraph{LM-REOS based models}
Using LM-REOS, we find solutions for $1.2 \leq \Ptrans\leq 5$~Mbars. For all of these models, 
$Z_1$ is nearly constant no matter what the values of $Z_2$ and $\Mcore$ are. This is because with 
the layer boundary so deep inside, $J_4$ is little sensitive to the mass distribution there.
As $Z_2$, in contrast, covers a wide range of 0--60\%, there are solutions with $Z_1=Z_2$,
unlike for Jupiter \citep{N+12}. 

For LM-REOS based models the layer boundary has to be put rather deep inside the planet in order to 
nearly suppress the influence of the matter interior to $\Ptrans$ to the values of $J_2$ and $J_4$. 
Otherwise, say for $\Ptrans < 1$~Mbar, the rise in $J_2$ could no longer be compensated for by a 
lower $Z_2$ value because $Z_2$ already goes to zero. This behavior is a direct consequence of the 
higher compressibility of the H EOS at sub-Mbar pressures (Figure \ref{fig:hugos}).

The total mass of heavy elements is 16--$20\:\ME$, less than predicted for Jupiter with LM-REOS
($\sim 30\ME$, \citealp{N+12}). However, Saturn's $M_Z$ corresponds to an overall  enrichment 
$Z_p=10$--15~\Zsol, higher than Jupiter's ($\sim 6\:\Zsol$).

\paragraph{Sesame based models}
Sesame EOS based models have the layer boundary between 0.2 and 2.5 Mbar. 
With $\Ptrans\to 0.2$~Mbar, the sensitivity of $J_4$ rises strongly 
(Figure \ref{fig:contribJ2n}) so that $Z_1$ has to decrease rapidly in order to not produce too high
$|J_4|$ values. There is no overlap of the functions $Z_1(\Ptrans)$ and $Z_2(\Ptrans)$, i.e.~Sesame 
EOS based Saturn models require a higher metallicity in the inner than in the outer envelope.
Because of the stiffness of the H-Sesame EOS up to 1 Mbar (Figure \ref{fig:hugos}), 
the layer boundary can be farther out and the $Z_1$ values as well as the total mass of heavy elements
can be up to 2.5 times higher than for LM-REOS based models. In fact, the layer boundary \emph{must} be 
farther out because the rise in $Z_2$ for increasing $\Ptrans$ is accompanied by a rapid decrease in $\Mcore$, 
which must not become negative, a direct effect of the higher compressibility of the H-Sesame EOS at higher 
pressures up to 10~Mbar (Figure \ref{fig:hugos}).

\paragraph{SCvH-i based models}
From the gross compressibility behavior of the Hugoniot up to 100~GPa one would expect the SCvH-i EOS based
solutions to fall between those for the other two EOSs. This in fact happens with respect to the parameters
$Z_2$, $\Mcore$, and $M_Z$. In particular, the just slightly higher compressibilities of SCvH-i compared to 
Sesame EOS below 100 GPa suggest just slighty \emph{lower} $Z_1$ values than for that EOS. However, 
the $Z_1$ values are relatively \emph{higher} than for the Sesame EOS based models. This can only be due to 
higher temperatures along the SCvH-i Saturn adiabat, as higher temperatures at given pressure level reduce 
the mass density, which allows for more heavy elements to be added. We will encounter the same argument 
again in Section \ref{sec:r_struc2}. In contrast to SG04, we do not find models with $Z_1=Z_2$. 
Presumably, this stems from the application of the more recent, accurate $J_{2n}$ data.

Both Sesame and SCvH-i EOS based models have $M_Z=25$--$30\:\ME$, a similar amount as Jupiter may have
($\sim 15$--$40\ME$; SG04). With $Z_p=17$--$21~\Zsol$, this is a larger enrichment 
than in comparable models for Jupiter (3--8~\Zsol). Note that SG04 use
an earlier value $\Zsol^{('89)}=0.019$.

\begin{figure}
\includegraphics[width=0.48\textwidth]{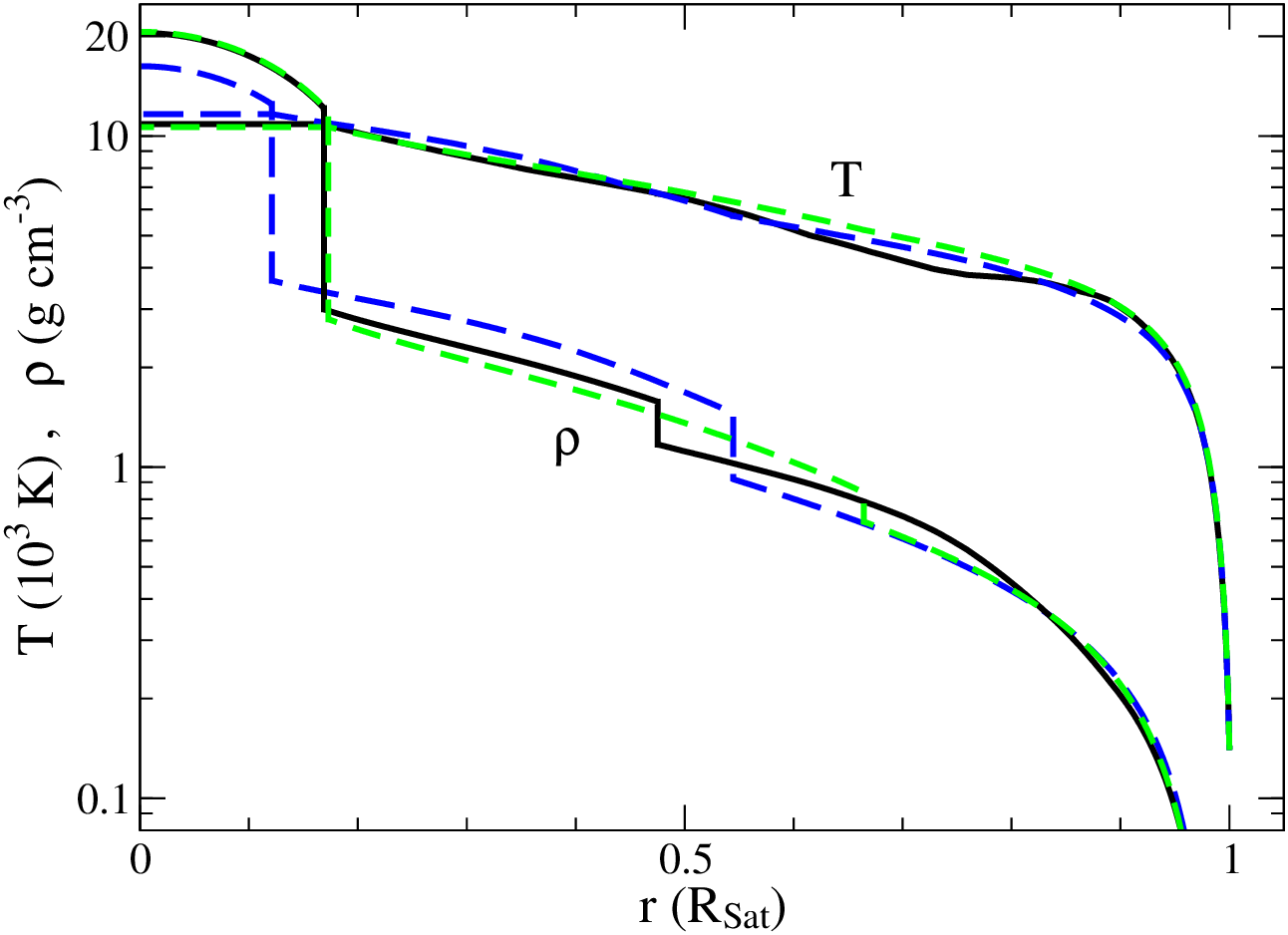}
\caption{\label{fig:profilesS} (Color online) 
Interior density and temperature profiles of the Saturn models that are highlighted in 
Fig.~\ref{fig:mcZZ_P12}c by filled circles.; \emph{solid, black}: model with LM-REOS, \emph{dashed, green}: 
SCvH-i, and \emph{long-dashed, blue}: with Sesame EOS. These models have an isothermal rock core. 
Between 0.95 and 1 \Rsat, the density changes by three orders of magnitude (not displayed), and between 
0.95 {\Rsat} and the boundary to the inner envelope, by one order of magnitude.}
\end{figure}

\paragraph{Internal profiles}
For each of the EOS, one selected interior profile is shown in Figure \ref{fig:profilesS}.
The temperatures along the SCvH-i Saturn adiabat in the outer envelope are indeed higher than for the 
Sesame adiabat. For the LM-REOS based adiabat, the onset of dissociation occurs at $\sim 0.7\Rsat$ where the 
temperature gradient flattens. A typical value of the density is 2 g~cm$^{-3}$ in the inner envelope, 
15 g~cm$^{-3}$ in a rocky core, and 8 g~cm$^{-3}$ in a water core, while in the outer envelope the 
density changes by four orders of magnitude.

Concluding, for each of the considered EOS, similar values of the metallicity in the outer envelope
(e.g., $3\times$ solar), in the inner envelope (e.g. 10--$30\times$ solar), and of the core 
(e.g., 0--$15\:\ME$ for rocky or 0--$18\:\ME$ for water cores) are possible. It depends mainly on the 
position of an internal compositional gradient (for instance in  form of a layer boundary), which
values are adopted. In contrast, the different EOSs require different locations of that gradient,
and LM-REOS predicts a lower total mass of heavy elements than the other two EOSs.

\subsection{LM-REOS based structure models with different helium abundances, rotation rates, and $J_4$ values} 
\label{sec:r_struc2}

The resulting atmospheric heavy element enrichment of $\sim 3\times$ solar of the LM-REOS based Saturn models 
is rather low compared to the measured carbon enrichment of $9\times$ solar. Therefore, we investigate in 
this Section qualitatively the effect of the assumed atmospheric helium abundance, of the rotation rate, 
and of the $J_4$ value on the resulting outer envelope metallicity.
As we are aiming to get higher $Z_1$ values than for the models in Section \ref{sec:r_struc1}, we only consider
higher $|J_4|$ values, up to the $6\sigma$ observational uncertainty (any $|J_{2n}|$ increases with the mass 
density in the sensitive region), lower $Y_1$ values ($Z_1$ decreases linearly with $Y_1$)
of respectively 0.16 and 0.1; but we consider a higher rotation rate 
(to face a possible reality), although it will require a reduction in the heavy element content of the outer 
part of a planet. For all of these models, the transition pressure is at 300 GPa and the core consists of rocks. 
At temperatures between 140 and 300 K, the analytic Van-der-Waals-gas EOS is used for H$_2$ and He, 
and the ideal gas EOS for molecular H$_2$O, but this has no relevant effect on the resulting adiabats.

\begin{figure}
\includegraphics[width=0.4\textwidth]{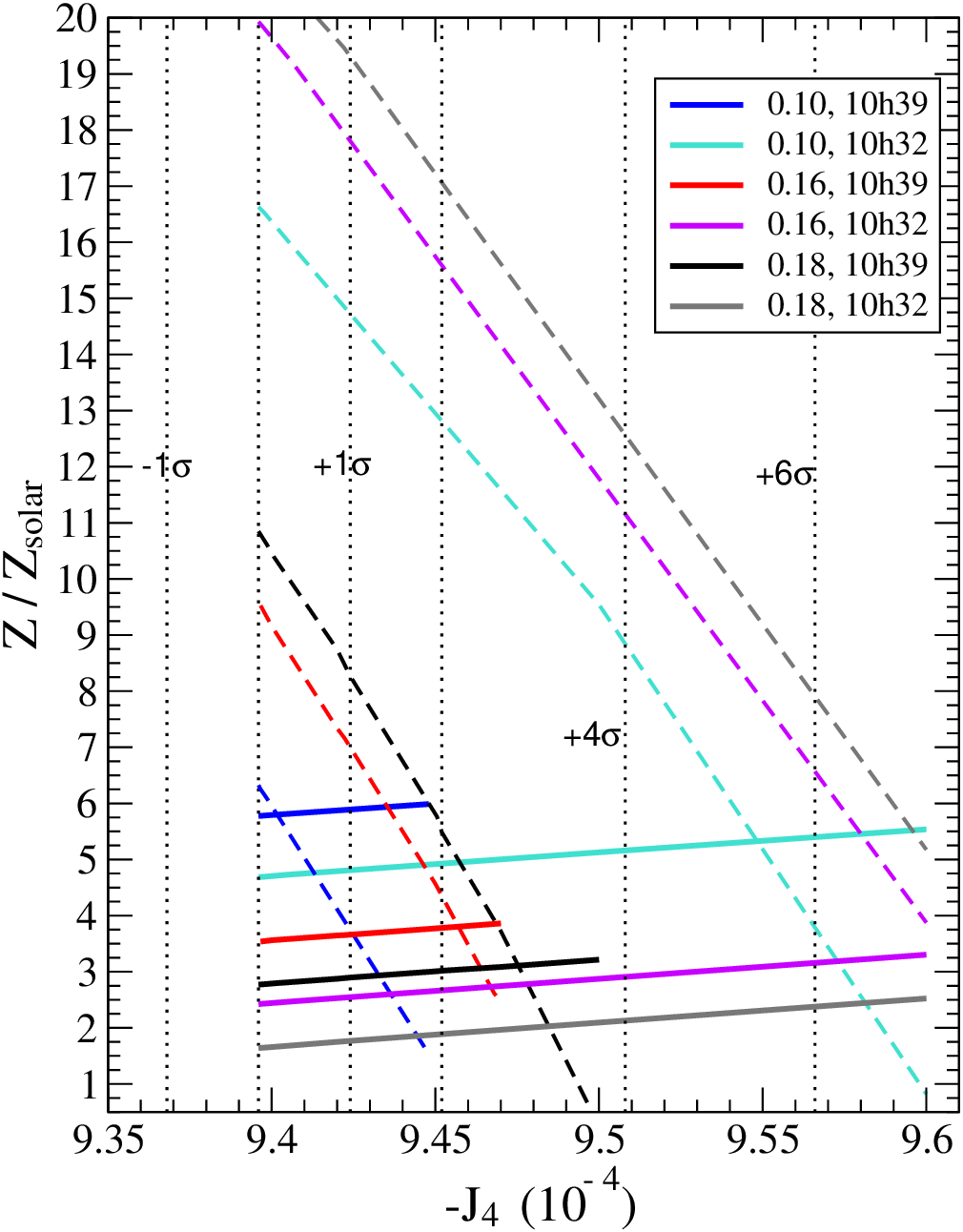}
\caption{\label{fig:ZZ_J4} Resulting outer (\emph{solid lines}) and inner envelope  metallicities 
(\emph{dashed lines}) in terms of the solar metallicity $Z_{\odot}=0.015$ for different He/(He+H) 
mass ratios of 0.10, 0.16, 0.18 and different periods of rotation of 10h 39m and 10h 32m as labeled,
and different given $J_4$ values up to the $6\sigma$ uncertainty (\emph{x-axis}). Some of the 
numerically possible models with $Z_2$ close to zero are not displayed. }
\end{figure}

Figure \ref{fig:ZZ_J4} shows the resulting enrichment factors for the metallicity in the outer 
envelope --to be compared with the (potentially) measured abundances-- and also in the inner envelope.

Obviously, the influence of $J_4$ on $Z_1$ is weak. Only for extremely high $|J_4|$ values of 
$9.6\times 10^{-4}$ the effect becomes of same size as the effect of a slower rotation by 7 minutes,
which is $\Delta Z_1\sim +1Z_{\odot}$.  
Note that $\Delta Z_1/Z_1 > 20\%$ constitutes a reather big amplification of the 1\% relative uncertainty 
in the rotation rate. 

In case a high rotation rate of 10h 32m should proof true, 
the heavy element enrichment would decrease down to $2\times$ solar for a helium abundance $\Yatm=0.18$. 
Note that we do not adjust the planet radius to the rotation rate. Interestingly, the rotation rate 
mainly affects the inner envelope metallicity. 
This can be explained by the response of $J_2$ on a change in $Z_1$ and $Z_2$ as described 
in \ref{apx:J2Z1Z2}.

The biggest effect on $Z_1$ can be achieved by lowering the helium abundance, where $\Delta Z_1\sim 1.5\%$ 
of heavy elements can be added for $\Delta Y_1\sim 2\%$ of helium. It is $|\Delta Z_1|/|\Delta Y_1| < 1$ 
because the less helium there is, the higher the specific heat of the material, and thus the
lower must be the temperature along the adiabat  to keep the entropy constant. Colder adiabats give
denser envelopes and thus allow for less heavy elements to be put into.

Despite the wide range of considered parameter values, with $Z_1\lesssim 6\ Z_{\odot}$ the possible 
atmospheric heavy element abundance remains clearly below the 9fold enrichment of carbon. Our 
LM-REOS based models therefore predict O/H to be less than $9\times$ solar.

\subsection{Moment of Inertia}\label{sec:r_nmoi}

\begin{figure}
\includegraphics[width=0.48\textwidth]{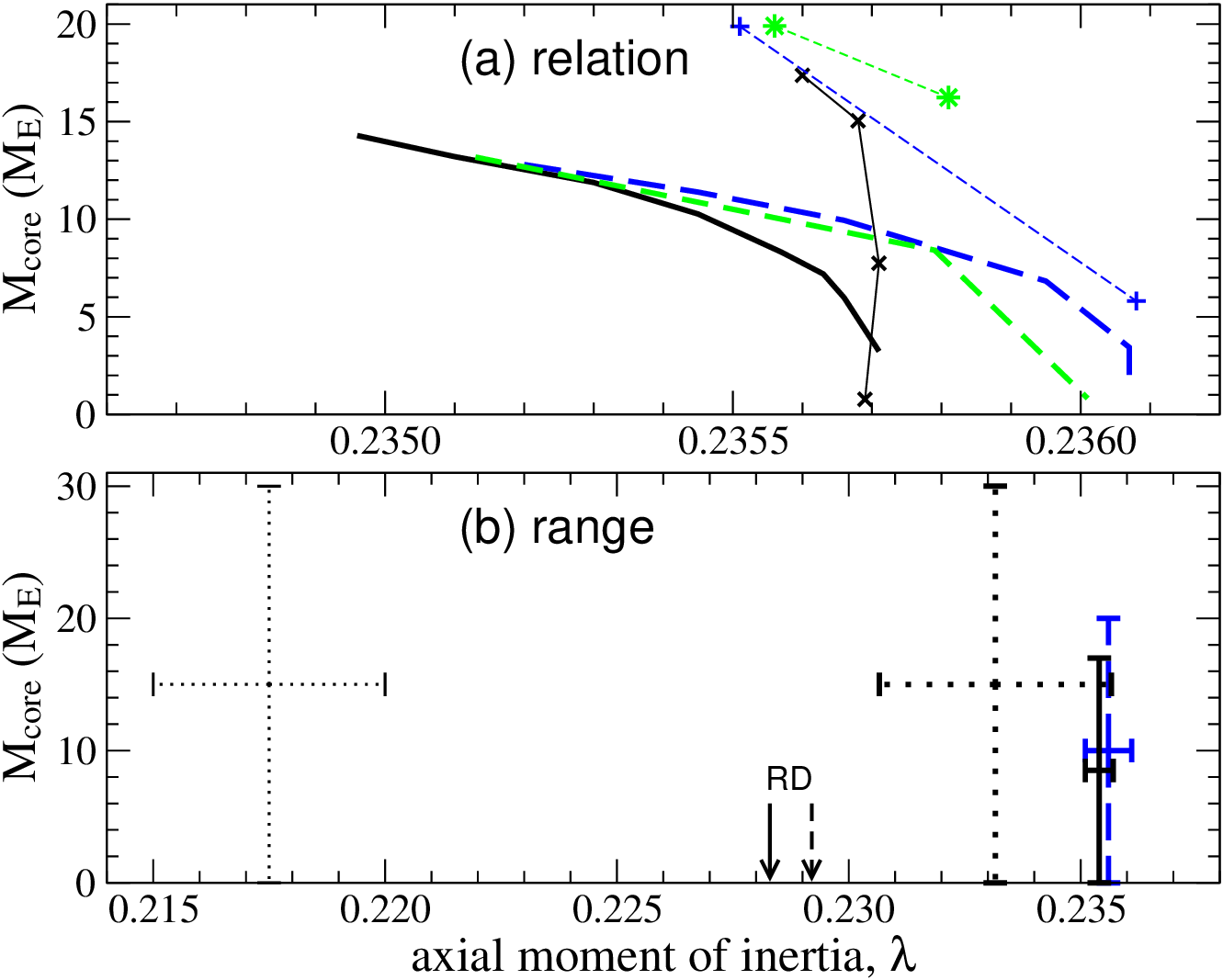}
\caption{\label{fig:nmoi} (Color online)
(a) Nondimensional axial moment of inertia---core mass relation for a subset of 
the structure models of Fig.~\ref{fig:mcZZ_P12} using the different EOSs LM-REOS (\emph{solid, black}), 
SCvH-i (\emph{dashed, green}), Sesame (\emph{long-dashed, blue}) with rock cores (\emph{lines}) or water cores 
(\emph{symbols}); 
(b)$^1$: Core mass range and $\lambda$-range; \emph{thin dotted:} from Ref.~\citet{Helled11} for a period
of rotation of 10h39m24s and scaling by $R_{\rm eq}^{-2}$ (R.~Helled, \emph{pers.~comm.~2013}); 
\emph{thick dotted:} same as thin dotted but scaled by $R_{\rm mean}^{-2}$;
\emph{Arrows} (vertical position has no meaning) Radau-Darwin approximation, 
see text for details.}
\end{figure}

\citet{Helled11} showed that an observational determination of Saturn's axial moment of inertia, $C$ 
would impose an additional constraint on Saturn's core mass, and that the necessary measurements can 
be provided by the \emph{Cassini} extended extended mission. 
Figure \ref{fig:nmoi} presents our results for Saturn's nondimensional moment of inertia, 
$\lambda=C/\Msat\Rmean^2$ for a representative subset of the models of Fig.~\ref{fig:mcZZ_P12}. 
As pointed out by and in agreement with \citet{Helled11}, we find different $\lambda$ values for different
core mass values for models that all meet the observed $J_2$ and $J_4$ values within their tight
$1\sigma$ bounds. In particular, for a fixed core EOS (rocks or water) and mantle EOS (SCvH, Sesame, 
or LM-REOS), and not too small core mass values ($\gtrsim 10\ME$), the relation between $\Mcore$ and 
$\lambda$ becomes unique. However, due to the uncertainties in the core and envelope EOS, a measurement 
of $\lambda$ could further, but not unambiguously constrain the core mass. 
Moreover, our physical EOS based, adiabatic models yield a very narrow possible range for $\lambda$, 
that even does not overlap with the prediction by \citet{Helled11}, see Figure \ref{fig:nmoi}b. 
Therefore, a measured $\lambda$ value would definitely be of great value for discriminating between 
competing Saturn models.

As moment of inertia measurements for gas giant planets are challenging, the Radau-Darwin approximation
is often used for an estimate. It expresses $\lambda$ in terms of the easier accessible $J_2$ and the 
expansion parameter $m=\omega^2 \Rmean^3/GM$, and becomes exact only in the limit of constant 
density bodies. 
Using $\Rmean=58201$~km, $P=10\rm h\,39m$, and the values of Table~\ref{tab:obs}, we calculate 
$\lambda^{(RD)}=0.2283$ (solid arrow in Figure \ref{fig:nmoi}b). The more consistent values of 
Table~1 in \citet{Helled11} suggest $\lambda^{(RD)}=0.2292$ (dashed arrow in Figure \ref{fig:nmoi}b). 
Interestingly, our interior model based values are $\sim 3\%$ higher than
$\lambda^{(RD)}$ 
indicating that higher-order 
deformations ($J_4$) play a non-negligible role in the internal mass distribution.
Note\footnote{modified after acceptance of this paper} that different scalings of $\lambda$ 
by either the equatorial radius as in \citet{Helled11} (R.~Helled, \emph{pers.comm.~2013}), or by the mean 
radius changes the value of Saturn's $\lambda$ by 7.2\%.

\subsection{Homogeneous Evolution}\label{sec:r_evol}


Homogeneous evolution implies a constant mean molecular weight in every mass shell of the planet with time, 
while inhomogeneous evolution also allows for vertical mass transport such as He rain or core erosion. 
In the considered case of homogeneous evolution, we get cooling times $\tau_{\rm Sat}$ of 2.56~Gyr for the 
LM-REOS model, 2.36~Gyr (SCvH-i EOS model), and 2.31~Gyr (Sesame EOS model). 

Neglecting the three contributions from angular momentum conservation, change of the energy of rotation, 
and from the time-dependence of the irradiation yields 0.05~Gyr longer cooling times for Saturn.
While with $\tau_{\rm Sat}\sim 2.5$~Gyr, the cooling time comes out significantly shorter than the age of 
the solar system of 4.56~Gyr ---commonly believed to also be the age of the planets within an uncertainty 
of a few Myr according to circumstellar disk observations \citep{Strom+93}--- 
it is in agreement with previous calculations. For instance, using the SCvH-i EOS, \citet{Guillot+95} 
compute $2.6\pm 0.2$~Gyr and \citet{FH03} 2.1~Gyr for an adiabatic, homogeneously evolving Saturn.

Our results show once more that the short cooling time of a homogeneously evolving Saturn is 
essentially independent on details of the model assumptions such as the size of the core. 
In other words, we confirm the well-known evidence for a real excess luminosity compared to the 
predicted luminosity from homogeneous evolution \citep{Pollack+77,Saumon+92,FH03}.

\section{Summary and Discussion}\label{sec:discussion}

\subsection{Gross features}

We have applied different EOSs to compute structure and evolution models for Saturn within the 
standard approach of a layered interior with only few layers that cool down homogeneously with time.
Because of the large applied input parameter space we have selected combinations of parameters that we 
believe yield a reliable estimate of the overall uncertainty in key internal structure parameters. In particular,
with LM-REOS and assuming $Y_1=0.18$ and $2\pi/\omega=$10h 39m we find $\Mcore = 0$--$17\ME$, $M_Z=16$--$20\ME$, 
$Z_{\rm atm}\lesssim 3\times$~solar, and $\tau_S=$2.6~Gyr; with Sesame-EOS we find $\Mcore=0$--$20\ME$, 
$Z_{\rm atm}\leq 7\times$~solar, $M_Z=26$--$30\ME$, and $\tau_S=$2.3~Gyr, while SCvH EOS based models have 
values in between. The value of $Z_{\rm atm}$ of the LM-REOS based models can be lifted up to a factor 
of two if $Y_1$ is lowered down 
to 0.10. With $\tau_S=2.3$--2.6 Gyr, the cooling time is significantly shorter than the age of the 
solar system, pointing to a failure of the cooling model that is to be sought beyond the uncertainties
in Saturn's composition or the EOS.

We encourage measurements of Saturn's moment of inertia and of the atmospheric abundance of helium and 
oxygen for discriminating between the wide range of possible Saturn models and for probing the 
underlying EOS in certain pressure ranges.

\subsection{O:H ratio}

\begin{figure}
\includegraphics[width=0.48\textwidth]{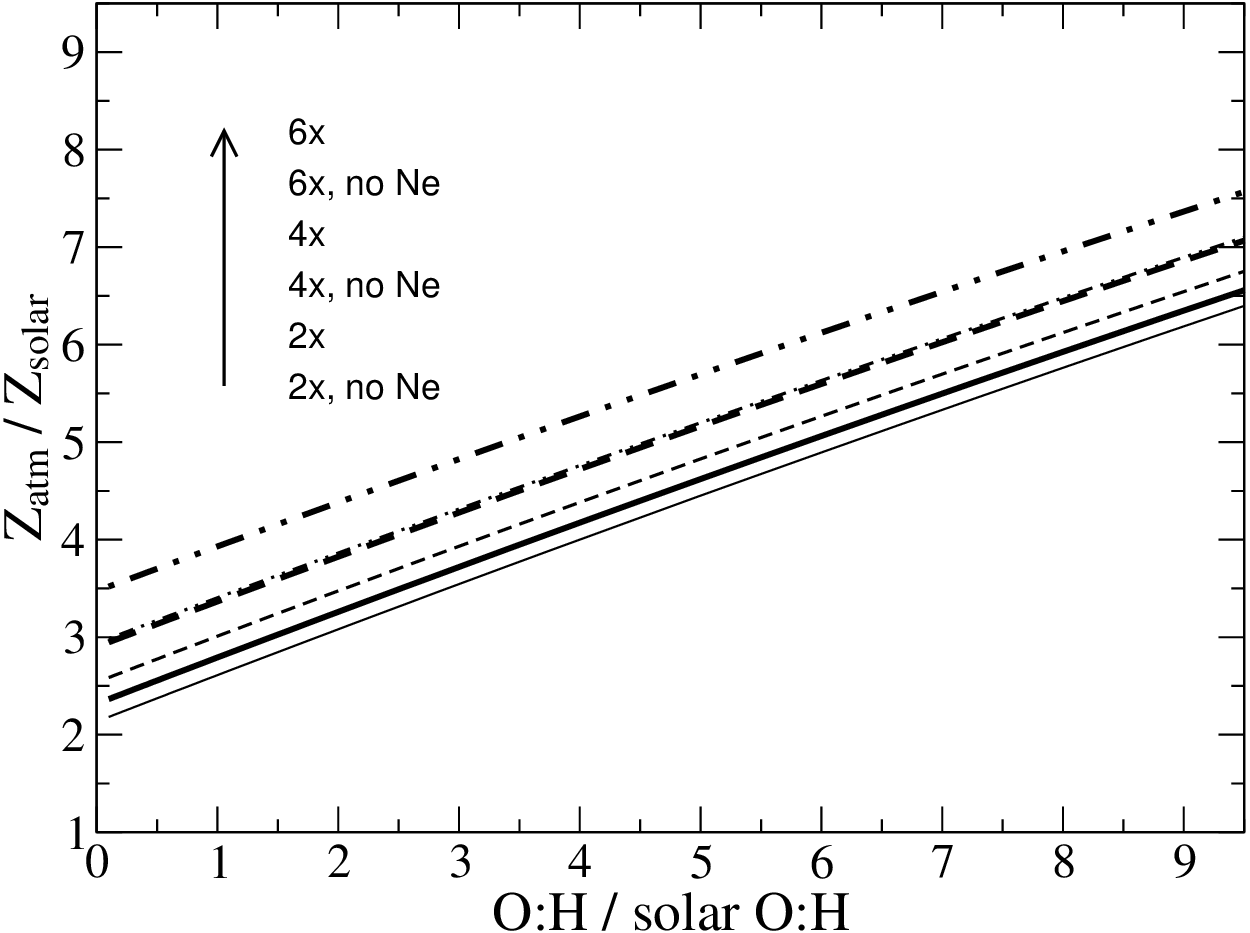}
\caption{\label{fig:ZOH}Relation between the atmosperic O:H ratio and the atmosheric metallicity, $\Zatm$ 
in solar units, for which different element abundances are assumed: elements apart from C and O are 2 times solar
(\emph{thick solid}), 4x solar (\emph{dashed}), and 6x solar (\emph{dot-dot-dashed}). \emph{Thin lines}: the
same, respectively, but with zero-Neon abundances. C:H is $9.12\times$ solar and O:H is displayed on the 
\emph{x-axis}.}
\end{figure}

The O:H ratio could in principle be derived from brightness temperature measurements 
at $\sim 1$~m wavelengths using LOFAR (D.~Gautier, \emph{pers.~comm.}), which is a set of ground-based 
radio antennas in western Europe. From a measured O:H, in addition to the already measured C:H, the 
atmospheric metallicity {\Zatm} can be estimated and compared with the $Z_1$ values of the Saturn 
structure models. For this purpose, we show possible O:H--{\Zatm} relations in Fig.~\ref{fig:ZOH}. 
For a given O:H, {\Zatm} has been computed as the sum of the heavy element particle abundances times 
their atomic weight, divided by the sum over all element abundances times their weight, where we assume 
He:H = 0.052 ($\Yatm=0.18$), C:H$=0.912\times$ solar, and $2\times$ (or $4\times$, $6\times$) 
solar abundances of the elements \{N, P, S, Ne, Ar, Kr, Xe, Mg, Al, Ca\}, using the solar system abundances
of \citet{Lodders03}. As Jupiter's atmosphere is observed to be strongly depleted in Ne, which may also 
be the case for Saturn if caused by He sedimentation \citep{WilMil10}, we also compute $\Zatm$ with  
Ne:H=0.  

Obviously, models with $Z_1=\sim 3\Zsol$ would imply a low O:H of only 2x solar; $Z_1\sim 6\times \Zsol$, 
the upper limit of the LM-REOS based  Saturn models, would imply O:H = 6--8$\times$ solar. Higher values 
are possible with the Sesame or the SCvH-i EOS. Concluding, a measured O:H could tremendously help to 
discriminate between the various Saturn models.

\subsection{Core mass and formation}\label{sec:d_mcoreform}

The possible core mass values of Jupiter and Saturn persistently attract attention, as the hope to 
infer the planet formation process from the ''face value'' of the present core mass continues to exist. 
Leading candidates for possible formation processes are the core accretion scenario, 
where the gaseous envelope is accreted onto a heavy-element core \citep{Alibert+05,DodRob+08,Kobayashi+12},
and the gravitational disk instability scenario, where the planet would form through self-contraction
of a gaseous cloud (\citealp[e.g.][]{HelledSchubert08}).

Let us assume the initial core mass of Saturn was $\gtrsim 15\ME$. What could that tells us?
Such a heavy core exceeds the maximum core mass of $8 \ME$ found by \citet{HelledSchubert08} and thus 
would rule out a disk-instability-kind formation for Saturn.
On the other hand, it would allow for a comfortably short time scale for core accretion formation
\citep{DodRob+08,Kobayashi+12}, even for a non-zero abundance of grains in the protoplanetary envelope \citep{DodRob+08}.

If the initial core mass of Saturn was $\sim 7\ME$, then both formation processes could have let
to the present Saturn. Given Saturn's high total heavy element mass of 16--30$\ME$, disk-instability 
formation would indicate a massive protosolar disk \citep{HelledSchubert09}, as well as an early
presence of planetesimals before formation was completed \citep{HelledSchubert08}. 

In the case of an initial core mass of $< 1\ME$, again both scenarios could be possible. 
Here, core accretion formation would require the absence of grains in the envelope in order to 
reduce the gas pressure induced by warm temperatures in an opaque medium \citep{HoriIkoma10}.

As main paths toward better constrained core mass properties, we support Helled's (2011) suggestion of a moment of 
inertia measurement, and emphasize the need for a better understanding of Saturn's envelope structure.

\subsection{Three-layer models and helium rain in Saturn}\label{sec:d_Herain}

The over and over confirmed finding of a too short cooling time for homogeneously evolving Saturn models,
regardless of details in the structure models, or of the underlying EOS used, and the simultaneously 
repeated mentioning of a high likelihood for an inhomogeneously evolving Saturn as a result of He 
sedimentation (e.g., \citealp{Pollack+77,SS77b,Saumon+92,Guillot+95,FH03}) cast doubt 
on the usefulness of Saturn models that ignore this process. 

The most recent theories of H/He phase separation predict demixing under relevant planetary
conditions to occur when hydrogen metallizes under high pressures while the helium atoms are still neutral
\citep{Morales+09,Lorenzen+09,Lorenzen+11}. In Jupiter, metallization occurs rather far out
in the planet at $\sim 0.9 R_J$ where the pressure is only 0.5~Mbar \citep{French+12}, but the lowest
pressure that could be achieved in H/He demixing calculations so far is 1~Mbar \citep{Lorenzen+11}.
Indeed, the ab initio data based H/He phase diagram suggests demixing in Saturn at least within 1--5 Mbar
\citep{Lorenzen+11}, which corresponds to the region beneath $\sim 2/3\: \Rsat$. 
Let us thus divide Saturn's mantle into three regions: an outer region down to 1~Mbar ($\sim 2/3\Rsat$),
an innermost helium-rich region where the sedimented helium dissolves again in its surrounding, 
possibly a helium-layer \citep{FH03}, and a middle region where the $P-T$ conditions favor 
H/He phase separation. 

In the middle region, the helium abundance at given $P,T$ will reduce until the remaining He atoms 
become miscible: the helium abundance will follow the demixing curve \citep{SS77b}. In $P$--$\rho$-space, 
such an inhomogeneous region could simply be described by a smoothed layer boundary. 
In $P$--$T$-space, the effect of an inhomogeneity could be significant, as it may be correlated with 
a strongly superadiabatic temperature gradient \citep{SS77b}, which would require higher envelope 
metallicities and affect the derived core mass \citep{LC12}.
Also important may be the induced differential rotation from sinking droplets that
conserve their angular momentum. \citet{Cao+12} showed  that a tiny differential rotation
would suffice to drive the magnetic field and explain its observed dipolarity.

In case the sedimented helium dissolves in the lower region, that part is well represented by an inner
envelope with enhanced abundance as in our models. But in case the helium rains down to the core,
a preferred solution to explain Saturn's high luminosity \citep{FH03}, the upper part of what we
count to be core material may in fact be helium. Thus, Saturn's maximum core mass could be lower 
than predicted by our models.

In the outer region we would mainly see the depletion in helium, because helium atoms from the upper
regions are transported repeatedly by convection down into the immiscibility region, where a fraction of 
them gets lost into the deep through sedimentation. Whatever happens to the compositional and temperature 
gradient deep in the planet, the outer, miscible envelope should remain homogeneous and adiabatic. 
Therefore, our results for the atmospheric helium and heavy element abundances are certainly caused or 
influenced by He-sedimentation, but are not expected to alter when He sedimentation would be explicitly 
accounted for.

\subsection{Why would we expect a (dis)continuous heavy element distribution?}

Standard three-layer models with heavy element discontinuity benefit from two additional parameters ($Z_2$, \Ptrans)
that can be used to fit the observational constraints. However, those models would even more benefit from 
a physical justification. 
At least, the assumption of a sharp layer boundary between two convective, homogeneous layers offers the advantage 
of a self-consistent picture, where upward particle transport across would be inhibited but not necessarily 
the heat transport. 
In contrast, a continuous, inhomogeneous heavy element distribution as suggested by \citet{LC12} might lead 
to a semi-convective boundary layer with reduced heat transport. Whether such a picture can explain the 
luminosities of both Jupiter and Saturn remains to be shown.

A possible origin for an inhomogeneous heavy element distribution could be the erosion of an initially big core,
with subsequent small-scale layer formation with small compositional gradients \citep{LC12}, and finally merging 
of the layers to a single one as seen in hydrodynamic simulations of fluids with both temperature and compositional gradients \citep{Wood+12}. 
Thus, a high, homogeneous metallicity in the inner envelope could be the result of an eroded core where the
core material went through stages of layer merging until the compositional gradient got large enough to stop
merging with what we now see as an outer envelope.

\subsection{Summarized main findings}\label{sec:d_summary}

\begin{itemize}
\item 
The H/He EOS strongly influences the atmospheric metallicity, $\Zatm$ and
the possible position of an internal layer boundary, but has little influence on the core mass and
the cooling time. We find $\Mcore=0$--$20\ME$ and $\tau_{\rm Sat}\sim 2.5$~Gyr.

\item 
The total mass of heavy elements in Saturn can be less (LM-REOS) or equal (Sesame, SCvH-i EOS) to that
in Jupiter, while the averaged enrichment is larger than that of comparable Jupiter models.

\item 
Our LM-REOS based Saturn models predict $\Zatm \sim 3\times$ solar ($\lesssim 6\times$) for an atmospheric 
helium abundance ${\Yatm}=18\%$  ($\Yatm=10\%$) by mass. The corresponding predicted 
maximum O:H ratio is $\sim 2 \times$ solar ($8\times$).

\item 
For Saturn, we calculate a non-dimensional axial moment of inertia $\lambda=0.235$ to 0.236.
\end{itemize}


We thank 
Daniel Gautier for illuminating conversations on abundance measurements, 
Johannes Wicht and Ravit Helled for interesting discussions,
Andreas Becker for computing the high-temperature extension of the H-REOS.2 Hugoniot curve,
and Winfried Lorenzen for discussions on H-He demixing.
The work presented in this paper is supported by the "Deutsche Forschungsgemeinschaft" (DFG) 
within the SFB 652 and the project RE 882/11.

\bibliographystyle{elsarticle-harv}
\bibliography{ms-refs}

\begin{thebibliography}{61}
\expandafter\ifx\csname natexlab\endcsname\relax\def\natexlab#1{#1}\fi
\expandafter\ifx\csname url\endcsname\relax
  \def\url#1{\texttt{#1}}\fi
\expandafter\ifx\csname urlprefix\endcsname\relax\def\urlprefix{URL }\fi

\bibitem[{Alibert et~al.(2005)Alibert, Mousis, Mordasini, and
  Benz}]{Alibert+05}
Alibert, Y., Mousis, O., Mordasini, C., Benz, W., 2005. {New Jupiter and Saturn
  formation models meet observations}. ApJ 626, L57.

\bibitem[{Anderson and Schubert(2007)}]{AS07}
Anderson, J., Schubert, G., 2007. {Saturn's Gravitational Field, Internal
  Rotation, and Interior Structure}. Science 317, 1384--1387.

\bibitem[{Atreya et~al.(2003)Atreya, Mahaffy, Niemann, Wong, and
  Owen}]{Atreya+03}
Atreya, S.~K., Mahaffy, P.~R., Niemann, H.~B., Wong, M.~H., Owen, T.~C., 2003.
  {Composition and origin of the atmosphere of Jupiter---an update, and
  implications for the extrasolar giant planets}. Planet. Space Sci. 51, 105.

\bibitem[{Boley et~al.(2011)Boley, Helled, and Payne}]{Boley+11}
Boley, A.~C., Helled, R., Payne, M.~J., 2011. {The heavy element composition of
  disk instability planets can range from sub- to super-nebular}. ApJ.

\bibitem[{Boriskov et~al.(2005)Boriskov, Bykov, Ilkaev, Selemir, Simakov,
  Trunin, Urlin, Shuikin, and Nellis}]{Boriskov+05}
Boriskov, G.~V., Bykov, A.~I., Ilkaev, R., Selemir, V.~D., Simakov, G.~V.,
  Trunin, R.~F., Urlin, V.~D., Shuikin, A.~N., Nellis, W.~J., 2005. {Shock
  compression of liquid deuterium up tp 109 GPa}. Phys. Rev. B 71, 092104.

\bibitem[{Campbell and Anderson(1989)}]{Campbell+89}
Campbell, J.~K., Anderson, J.~D., 1989. {Gravity field of the Saturnian system
  from Pioneer and Voyager tracking data.} Astronom. J. 97, 1485.

\bibitem[{Cao et~al.(2012)Cao, Russell, Wicht, Christensen, and
  Dougherty}]{Cao+12}
Cao, H., Russell, C.~T., Wicht, J., Christensen, U.~R., Dougherty, M.~K., 2012.
  {Saturn's high-degree magnetic moments: Evidence for a unique planetary
  dynamo}. Icarus 221, 388--394.

\bibitem[{Conrath and Gautier(2000)}]{CG00}
Conrath, B., Gautier, D., 2000. {Saturn Helium Abundance: A Reanalysis of
  Voyager Measurements}. Icarus 144, 124.

\bibitem[{Conrath et~al.(1984)Conrath, Gautier, Hanel, and
  Hornstein}]{Conrath+84}
Conrath, B.~J., Gautier, D., Hanel, R.~A., Hornstein, J.~S., 1984. {The helium
  abundance of Saturn from Voyager measurements}. Astrophys. J 282, 807--815.

\bibitem[{Desch and Kaiser(1981)}]{DK81}
Desch, M.~D., Kaiser, M.~L., 1981. {Voyager measurement of the rotation period
  of Saturn's magnetic field}. Geophys. Res. Lett. 8, 253.

\bibitem[{Dodson-Robinson et~al.(2010)Dodson-Robinson, Bodenheimer, Laughlin,
  Willacy, Turner, and Beichman}]{DodRob+08}
Dodson-Robinson, S., Bodenheimer, P., Laughlin, G., Willacy, K., Turner, N.~J.,
  Beichman, C.~A., 2010. {Saturn forms by core accretion in 3.4 Myr}. ApJ 688,
  L99.

\bibitem[{Fletcher et~al.(2009)Fletcher, Orton, Teanby, Irwin, and
  Bjoraker}]{Fletcher+09}
Fletcher, L.~N., Orton, G., Teanby, N., Irwin, P., Bjoraker, G., 2009. {Methane
  and its isotopologues on Saturn from Cassini/CIRS observations}. Icarus 199,
  351--367.

\bibitem[{Fortney and Hubbard(2003)}]{FH03}
Fortney, J.~J., Hubbard, W.~B., 2003. {Phase Separation in Giant Planets:
  Inhomogeneous evolution of Saturn}. Icarus 164, 228.

\bibitem[{Fortney et~al.(2011)Fortney, Ikoma, Nettelmann, Guillot, and
  Marley}]{Fortney+11}
Fortney, J.~J., Ikoma, M., Nettelmann, N., Guillot, T., Marley, M.~S., 2011.
  {Self-consistent Model Atmospheres and the Cooling of the Solar System Giant
  Planets}. ApJ 729, 32.

\bibitem[{French et~al.(2012)French, Becker, Lorenzen, Nettelmann,
  Bethkenhagen, Wicht, and Redmer}]{French+12}
French, M., Becker, A., Lorenzen, W., Nettelmann, N., Bethkenhagen, M., Wicht,
  J., Redmer, R., 2012. {Ab initio simulations for material properties along
  the Juptier adiabat}. ApJS 202, A5.

\bibitem[{French et~al.(2009)French, Mattsson, Nettelmann, and
  Redmer}]{French+09}
French, M., Mattsson, T.~R., Nettelmann, N., Redmer, R., 2009. {Equation of
  state and phase diagram of water at ultrahigh pressures as in planetary
  interiors}. Phys. Rev. B 79, 054107.

\bibitem[{Guillot(1999)}]{Gui99}
Guillot, T., 1999. {A comparison of the interiors of Jupiter and Saturn}.
  Planet. Space Sci. 47, 1183.

\bibitem[{Guillot et~al.(1995)Guillot, Chabrier, Gautier, and
  Morel}]{Guillot+95}
Guillot, T., Chabrier, G., Gautier, D., Morel, P., 1995. {Effect of radiative
  transport on the evolution of Jupiter and Saturn}. ApJ 450, 463.

\bibitem[{Guillot and Gautier(2007)}]{GuiGau07}
Guillot, T., Gautier, D., 2007. {The Giant Planets}. In: Schubert, G., Spohn,
  T. (Eds.), Treatise of Geophysics, vol.~10, Planets and Moons. Amsterdam:
  Elsevier, p. 439 (arXiv:0912:2019).

\bibitem[{Gurnett et~al.(2007)Gurnett, Persoon, Kurth, Groene, Averkamp, and
  Dougherty}]{Gurnett+07}
Gurnett, D.~A., Persoon, A.~M., Kurth, W.~S., Groene, J.~B., Averkamp, T.~F.,
  Dougherty, M. K.~andSouthwood, D.~J., 2007. {The Variable Rotation Period of
  the Inner Region of Saturn's Plasma Disk}. Science 316, 442--445.

\bibitem[{Helled(2011)}]{Helled11}
Helled, R., 2011. {Constraining Saturn's core properties by a measurement of
  ots moment of inertia-implications to the Cassini Solstice Mission}. ApJ 735,
  L16.

\bibitem[{Helled et~al.(2010)Helled, Bodenheimer, and Lissauer}]{Helled+10}
Helled, R., Bodenheimer, P., Lissauer, J.~J., 2010. {Composition of massive
  planets}. Proceeding IAU symp. 276, 119.

\bibitem[{Helled and Schubert(2008)}]{HelledSchubert08}
Helled, R., Schubert, G., 2008. {Core formation in gaseous protoplanets}.
  Icarus 0, 0.

\bibitem[{Helled and Schubert(2009)}]{HelledSchubert09}
Helled, R., Schubert, G., 2009. {Heavy element enrichement of a Jupiter-mass
  protoplanet as a function of orbital distance}. ApJ 697, 1256.

\bibitem[{Helled et~al.(2009{\natexlab{a}})Helled, Schubert, and
  Anderson}]{Helled+09a}
Helled, R., Schubert, G., Anderson, J.~D., 2009{\natexlab{a}}. {Empirical
  models of pressure and density in Saturns interior: Implications for the
  helium concentration, its depth dependence, and Saturns precession rate}.
  Icaus 199, 368--377.

\bibitem[{Helled et~al.(2009{\natexlab{b}})Helled, Schubert, and
  Anderson}]{Helled+09b}
Helled, R., Schubert, G., Anderson, J.~D., 2009{\natexlab{b}}. {Jupiter and
  Saturn Rotation Periods}. Planet. Space Sci. 57, 1467--1473.

\bibitem[{Holst et~al.(2012)Holst, Redmer, Gryaznov, Fortov, and
  Iosilevskiy}]{Holst+12}
Holst, B., Redmer, R., Gryaznov, V.~K., Fortov, V.~E., Iosilevskiy, I.~L.,
  2012. {Hydrogen and helium in shock wave experiments, ab initio simulations
  and chemical picture modeling}. Eur.~Phys.~J.~D 66, 104.

\bibitem[{Hori and Ikoma(2010)}]{HoriIkoma10}
Hori, Y., Ikoma, M., 2010. {Critical core masses for gas giant formation with
  grain free-envelopes}. ApJ 714, 1343.

\bibitem[{Hubbard(1999)}]{Hubbard99}
Hubbard, W.~B., 1999. {Gravitational Signature of Jupiter's Deep Zonal Flows}.
  Icarus 137, 357--359.

\bibitem[{Hubbard and Marley(1989)}]{HM89}
Hubbard, W.~B., Marley, M.~S., 1989. {Optimized Jupiter, Saturn, and Uranus
  interior models}. Icarus 78, 102.

\bibitem[{Jacobson et~al.(2006)Jacobson, Antresian, Bordi, Criddle, Ionasescu,
  Jones, Mackenzie, Meek, Parcher, and Pelletier}]{J+06}
Jacobson, R.~A., Antresian, P.~G., Bordi, J.~J., Criddle, K.~E., Ionasescu, R.,
  Jones, J.~B., Mackenzie, R.~A., Meek, M.~C., Parcher, D., Pelletier, F.~J.,
  2006. {The gravity field of the Saturnian system from satellite observations
  and spacecraft tracking data132}. Astronom. J. 132, 2520.

\bibitem[{Kerley(2003)}]{Kerley03}
Kerley, G., 2003. Equations of state for hydrogen and deuterium. Tech. rep.,
  SANDIA REPORT, SAND2003-3613.

\bibitem[{Kerley(2004a)}]{Kerley04a}
Kerley, G., 2004a. {Structures of the planets Jupiter and Saturn}. Tech. rep.,
  Kerley Tech. Services, Report KTS04-1.

\bibitem[{Kerley(2004b)}]{Kerley04b}
Kerley, G., 2004b. {An Equation of State for Helium}. Tech. rep., Kerley Tech.
  Services, Report KTS04-2.

\bibitem[{Knudson and Desjarlais(2009)}]{KnudDesj09}
Knudson, M.~D., Desjarlais, M.~P., 2009. {Shock Compression of Quartz to 1.6
  TPa: Redefining a Pressure Standard}. Phys. Rev. Lett. 103, 225501.

\bibitem[{Knudson et~al.(2004)Knudson, Hanson, Bailey, Hall, Asay, and
  Deeney}]{Knudson+04}
Knudson, M.~D., Hanson, D.~L., Bailey, J.~E., Hall, C.~A., Asay, J.~R., Deeney,
  C., 2004. {Principal Hugoniot, Reveberating Wave, and mechanical reshock
  measurement of liquid deuterium to 400 GPa using plate impact techniques.}
  Phys. Rev. B 69, 144209.

\bibitem[{Kobayashi et~al.(2012)Kobayashi, Ormel, and Ida}]{Kobayashi+12}
Kobayashi, H., Ormel, C.~W., Ida, S., 2012. {Rapid formation of Saturn after
  Jupiter completion}. ApJ submitted.

\bibitem[{Leconte and Chabrier(2012)}]{LC12}
Leconte, J., Chabrier, G., 2012. A new vision on giant planet interiors: the
  impact of double-diffusive convection. A\&A 540, A20.

\bibitem[{Lindal et~al.(1985)Lindal, Sweetnam, and Eshleman}]{Lindal+85}
Lindal, G., Sweetnam, D.~N., Eshleman, V.~R., 1985. {The atmosphere of Saturn:
  an analysis of the Voyager occultation measurements}. Astronom. J. 90, 1136.

\bibitem[{Lodders(2003)}]{Lodders03}
Lodders, K., 2003. {Solar System Abundances and Condensation Temperatures of
  the Elements.} ApJ 591, 1220.

\bibitem[{Lorenzen et~al.(2009)Lorenzen, Holst, and Redmer}]{Lorenzen+09}
Lorenzen, W., Holst, B., Redmer, R., 2009. {Demixing of Hydrogen and Helium at
  Megabar Pressures}. Phys. Rev. Lett. 102, 5701.

\bibitem[{Lorenzen et~al.(2011)Lorenzen, Holst, and Redmer}]{Lorenzen+11}
Lorenzen, W., Holst, B., Redmer, R., 2011. {Metallization in hydrogen-helium
  mixtures}. Physical Review B 84, 235109.

\bibitem[{Lyon and Johnson(1992)}]{SESAME}
Lyon, S., Johnson, J. D.~e., 1992. Sesame: Los alamos national laboratory
  equation of state database. Tech. rep., LANL report no. LA-UR-92-3407.

\bibitem[{Miller and Fortney(2011)}]{Miller+11}
Miller, N., Fortney, J.~J., 2011. {The heavy-element masses of extrasolar giant
  planets, revealed}. ApJ 736, L29.

\bibitem[{Mizuno(1980)}]{Mizuno80}
Mizuno, H., 1980. Formation of the giant planets. Prog.~Theo.~Phys. 64, 544.

\bibitem[{Morales et~al.(2009)Morales, Schwegler, Ceperley, Pierleoni, Hamel,
  and Caspersen}]{Morales+09}
Morales, M.~A., Schwegler, E., Ceperley, D., Pierleoni, C., Hamel, S.,
  Caspersen, K., 2009. {Phase separation in hydrogen-helium mixtures at Mbar
  pressures}. PNAS 106, 1324--1329.

\bibitem[{Nellis et~al.(1983)Nellis, Mitchell, van Thiel, Devine, Trainor, and
  Brown}]{Nellis+83}
Nellis, W.~J., Mitchell, A.~C., van Thiel, M., Devine, G.~J., Trainor, R.~J.,
  Brown, N., 1983. {EOS data for molecular H and D at shock pressures in the
  range 2-76 GPa}. J.~Chem.~Phys. 79, 1480.

\bibitem[{Nettelmann et~al.(2012)Nettelmann, Becker, Holst, and Redmer}]{N+12}
Nettelmann, N., Becker, A., Holst, B., Redmer, R., 2012. {Jupiter models with
  improved hydrogen EOS (H-REOS.2)}. ApJ 750, A52.

\bibitem[{Nettelmann et~al.(2011)Nettelmann, Fortney, Kramm, and Redmer}]{N+11}
Nettelmann, N., Fortney, J., Kramm, U., Redmer, R., 2011. {Thermal evolution
  and structure models of the transiting super-Earth GJ1214b}. ApJ 750, A52.

\bibitem[{Nettelmann et~al.(2008)Nettelmann, Holst, Kietzmann, French, Redmer,
  and Blaschke}]{N+08}
Nettelmann, N., Holst, B., Kietzmann, A., French, M., Redmer, R., Blaschke, D.,
  2008. {Ab initio equation of state data for hydrogen, helium, and water and
  the internal structure of Jupiter}. ApJ 683, 1217.

\bibitem[{Orton and Ingersoll(1980)}]{OrtonIng80}
Orton, G.~S., Ingersoll, A.~P., 1980. {Saturn's atmospheric temperature
  structure and heat budget}. J. Geophys. Res. 85, 5871.

\bibitem[{Pollack et~al.(1977)Pollack, Grossman, Moore, and
  Graboske}]{Pollack+77}
Pollack, J.~B., Grossman, A.~S., Moore, R., Graboske, H.~C., 1977. {A
  calculation of Saturn's gravitational contraction history}. Icarus 30,
  111--128.

\bibitem[{Saumon et~al.(1995)Saumon, Chabrier, and "van Horn"}]{SCvH95}
Saumon, D., Chabrier, G., "van Horn", H.~M., 1995. {An equation of state for
  low-mass stars and giant planets}. ApJS 99, 713.

\bibitem[{Saumon and Guillot(2004)}]{SG04}
Saumon, D., Guillot, T., 2004. {Shock Compression of Deuterium and the
  Interiors of Jupiter and Saturn}. ApJ 609, 1170.

\bibitem[{Saumon et~al.(1992)Saumon, Hubbard, Chabrier, and van
  Horn}]{Saumon+92}
Saumon, D., Hubbard, W.~B., Chabrier, G., van Horn, H.~M., 1992. {The role of
  the molecular-metallic transition of hydrogen in the evolution of Jupiter,
  Saturn, and brown dwarfs.} ApJ 391, 827--831.

\bibitem[{Stevenson and Salpeter(1977)}]{SS77b}
Stevenson, D.~J., Salpeter, E.~E., 1977. {The dynamics and helium distribution
  in hydrogen-helium fluid planets}. ApJS 35, 239--261.

\bibitem[{Strom et~al.(1993)Strom, Edwards, and Skrutskie}]{Strom+93}
Strom, S.~E., Edwards, S., Skrutskie, M.~F., 1993. {Evolutionary time scales
  for circumstellar disks associated with intermediate- and solar-type stars}.
  In: Levy, E.~H., Lunine, J.~I. (Eds.), {Protostars and Planets III}.
  Amsterdam: Elsevier, pp. 837--866.

\bibitem[{von Zahn et~al.(1998)von Zahn, Hunten, and Lehmacher}]{Zahn+98}
von Zahn, U., Hunten, D.~M., Lehmacher, G., 1998. {Helium in Jupiter's
  atmosphere: Results from the Galileo probe helium interferometer experiment}.
  J. Geophys. Res. 103, 22815.

\bibitem[{Wilson and Militzer(2010)}]{WilMil10}
Wilson, H.~F., Militzer, B., 2010. {Sequestration of noble gases in giant
  planet interiors}. Phys.~Rev.~Lett. 104, 121101.

\bibitem[{Wood et~al.(2012)Wood, Garaud, and Stellmach}]{Wood+12}
Wood, T.~S., Garaud, P., Stellmach, S., 2012. {A new model for mixing by
  double-diffusive convection (semi-convection). II. The transport of heat and
  composition through layers}. arXiv:1212.1218v1 [astro-ph].

\bibitem[{Zharkov and Trubitsyn(1978)}]{ZT78}
Zharkov, V.~N., Trubitsyn, V.~P., 1978. Physics of Planetary Interiors. Tucson,
  AZ: Parchart.

\end{thebibliography}

\appendix
\section{Erratum Jupiter-paper}

In the Jupiter-II paper by \citet{N+12} on Jupiter structure and homogeneous evolution models it was stated
that including the energy of rotation in the thermal evolution leads to a reduced luminosity with time
and thus a \emph{longer} cooling time. The latter statement must be corrected for a \emph{shorter} cooling
time: the energy of rotation will \emph{not} be released by radiation at a later time. Therefore, Jupiter's
cooling time is \emph{decreased} by 0.2~Gyr (and not prolonged by 0.2~Gyr as stated in that paper).
The given final value for cooling time of 4.41~Gyr, when in addition the time-dependence of $L_{\odot}$ is
included, remains valid.

\section{Estimate $\Delta Z_2$ as a function of $\Delta Z_1$} \label{apx:J2Z1Z2}

We here derive an estimate for the  necessary change in $Z_2$ in response to a change $\Delta Z_1$ when
$J_2$ is to be kept unchanged. According to Equation \ref{eq:J2n} we can approximate $J_2$ by 
\begin{equation}\label{eq:J2approx}
J_2\approx V_{L1}\,\bar{\rho}_{1}\, r^4(\bar{\rho}_1) + V_{L2}\,\bar{\rho}_{2}\,r^4(\bar{\rho}_2),
\end{equation}
where $\bar{\rho}_i$, $i=1,2$, is the mean density of layer No.~$i$, and $V_{Li}$ its volume. 
The contribution from the small, central core is neglected. According to the additive 
volume law for mixtures we can write $\rho^{-1}=(1-Z)\rho_{\rm H,He}^{-1} + Z\rho_Z^{-1}$ and thus
\begin{equation}\label{eq:drhodZ}
	\frac{d\rho}{dZ} = \rho^2\left(\rho^{-1}_{\rm H,He} - \rho_Z^{-1}\right)\quad,
\end{equation}
so that $\Delta \rho_1\sim \bar{\rho}_1^2\,\Delta Z_1$ and $\Delta \rho_2\sim \bar{\rho}_2^2\,\Delta Z_2$.
Because the value of $J_2$ is an observational constraint, $\Delta J_2$ must be zero under the perturbations
$\Delta \bar{\rho}_1$ and $\Delta \bar{\rho}_2$. With Equations (\ref{eq:J2approx}) and (\ref{eq:drhodZ}) 
we thus have
\begin{equation}
	0 = \Delta J_2 \approx V_{L1}\:\Delta Z_1\: \bar{\rho}^2_1\: r^4(\bar{\rho}_1) 
	+ V_{L2}\:\Delta Z_2\: \bar{\rho}_2^2 \: r^4(\bar{\rho}_2)\quad, 
\end{equation}
hence
\begin{equation}\label{eq:factors}
	\Delta Z_2 = -\Delta Z_1 \left(\frac{V_{L1}}{V_{L2}}\right)\left(\frac{r(\bar{\rho}_1)}{r(\bar{\rho}_2)}\right)^4
	\left(\frac{\bar{\rho_1}}{\bar{\rho}_2}\right)^2\:.
\end{equation}
For a typical three-layer Saturn model as shown in Figure~\ref{fig:profilesS}, 
$r(\Ptrans)\sim 0.5 \Rsat$, leading to $V_{L1}/V_{L2}\sim 1/0.5^3= 2^3$, 
$r(\bar{\rho}_1)=0.7\Rsat$, $r(\bar{\rho}_2)=0.35 \Rsat$, 
leading to $\left(r(\bar{\rho}_1)/r(\bar{\rho}_2)\right)^4\sim 2^4$,
$\bar{\rho}_1\sim 0.7\:\rm g\,cm^{-3}$, $\bar{\rho}_1\sim 2.05\:\rm g\,cm^{-3}$ 
leading to $\left(\bar{\rho}_1/\bar{\rho}_2\right)^2 \sim (1/3)^2$, and thus
$\Delta Z_2/\Delta Z_1 \approx -2^7/3^2 \approx -16$, in reasonable 
agreement with $\Delta Z_2/\Delta Z_1 \approx -10$ as seen in Figure \ref{fig:ZZ_J4}.

\end{document}